\documentstyle[12pt,aasms4]{article}

\begin{document}

\title{Stellar Abundances in the Early Galaxy and Two $r$-Process Components}
\author{Y.-Z. Qian\altaffilmark{1} and G. J. Wasserburg\altaffilmark{2}}
\altaffiltext{1}{T-5, MS B283, Theoretical Division, Los Alamos National
Laboratory, Los Alamos, NM 87545; qian@paths.lanl.gov.}
\altaffiltext{2}{The Lunatic Asylum,
Division of Geological and Planetary Sciences, California
Institute of Technology, Pasadena, CA 91125.}

\begin{abstract}
We present quantitative predictions for the abundances of 
$r$-process elements
in stars formed very early in the Galactic history
using a phenomenological two-component $r$-process model
based on the $^{129}$I and $^{182}$Hf inventory 
in the early solar system. This model assumes
that a standard mass of the ISM dilutes the debris from an
individual supernova.
High frequency supernova H events and low frequency supernova L
events are proposed in the model with characteristics determined by
the meteoritic data on $^{129}$I and $^{182}$Hf. 
The yields in an H or L event are obtained from these characteristics and
the solar $r$-process abundances under the assumption that
the yield template for the high mass ($A > 130$) nuclei 
associated with $^{182}$W or the low mass ($A\leq 130$) nuclei
associated with $^{127}$I is the same for both the H and L events
and follows the corresponding solar $r$-pattern in each mass region.
This choice of the yield templates is justified by the regular
solar-like $r$-process abundance pattern for Ba and higher atomic numbers
observed in very metal-poor stars. 

The abundance of Eu, not Fe, is 
proposed as a key guide to the age of very metal-poor stars. 
We predict that stars with $\log\epsilon({\rm Eu})=-2.98$ to $-2.22$ were
formed from an ISM
contaminated most likely by a single H event within the first $\sim 10^7$~yr
of the Galactic history and should have an Ag/Eu abundance ratio less than
the corresponding solar $r$-process value by a factor of at least 10. 
Many of the very metal-poor stars observed so far are
considered here to have been formed from an ISM contaminated by many 
($\sim 10$) $r$-process events. Stars formed from an ISM  
contaminated only by a pure L event would have an Ag/Eu ratio
higher than the corresponding solar $r$-process value but 
would be difficult to find due to the low frequency of the L events. 
However, variations in the relative abundances of 
the low and high mass regions should be detectable in very metal-poor stars.

\end{abstract}


\section{Introduction}

In this paper, we present 
the estimated abundances of elements produced by two
distinct types of $r$-process events based on a phenomenological model for
the production of the relevant nuclei. 
There are ongoing active investigations 
determining elemental abundances 
in very metal-poor stars with particular emphasis on the 
abundances of Th and Eu which are important for Galactic chronometry
(cf. Cowan et al. 1997, 1999). 
These studies also address  
the question of the universality of the solar system ``$r$-process'' 
abundance pattern (hereafter referred to as the solar $r$-pattern,
cf. Cowan et al. 1995, 1996; Sneden et al. 1994, 1996, 1998, 1999;
Crawford et al. 1998). In particular, Sneden et al. (1996, 1998, 1999)
have demonstrated that the abundances of elements in the 
Pt peak (at mass number $A\sim 195$) and down to Ba ($A\sim 135$)
in CS 22892--052 ([Fe/H] $=-3.1$),
HD 115444 ([Fe/H] $=-2.77$), and HD 126238 ([Fe/H] $=-1.67$)
are in remarkable accord with the solar $r$-pattern.

It has generally been considered that the abundances of the 
$r$-process nuclei are due to a single generic type of events
because the $r$-pattern observed
in very metal-poor stars agrees so well with that of the solar system.
Two classes of $r$-process calculations have been 
carried out: both rely on theoretical studies of properties of very
neutron-rich nuclei, but one uses simple parametrizations of the
astrophysical 
environment (e.g., Kratz et al. 1993) while the other has a detailed
astrophysical context (e.g., Meyer et al. 1992; 
Takahashi, Witti, \& Janka 1994; Woosley et al. 1994).
So far, no stellar model succeeds in generating the conditions required to
produce the entire solar $r$-pattern (Witti, Janka, \& Takahashi 1994; 
Qian \& Woosley 1996; Hoffman, Woosley, \& Qian 1997; Meyer \& Brown 1997;
Freiburghaus et al. 1999). 

However, it has been shown that the meteoritic data on
the inventory of $^{129}$I
(with a lifetime of $\bar\tau_{129}=2.27\times 10^7$~yr) and $^{182}$Hf 
($\bar\tau_{182}=1.30\times 10^7$~yr) in the early solar system
require that the stellar sources responsible for 
these two nuclei be decoupled (Wasserburg, Busso, \& Gallino 1996, 
hereafter WBG96). 
The nuclide $^{129}$I is a pure $r$-process 
product (Cameron 1993). As $^{182}$Hf cannot be produced effectively 
by the $s$-process in Asymptotic Giant Branch stars 
(Wasserburg et al. 1994; Busso, Gallino, \& Wasserburg 1999),
the solar inventory of $^{182}$Hf must also have originated from 
an $r$-process. The meteoritic requirement for diverse $r$-process events
can be seen from the following argument. 

Let us first assume that there
were only a single type of $r$-process events.
With the recognition that both $^{127}$I and $^{129}$I must be produced 
concurrently and at about the same yields, the observed abundance ratio
$(^{129}{\rm I}/{^{127}{\rm I}})_{\rm SSF}=10^{-4}$ 
(Reynolds 1960; Jeffrey \& Reynolds 1961; see Brazzle et al. 1999 
for a recent summary) at the time of solar 
system formation (SSF) then demands that 
the last injection of $^{129}$I into the interstellar medium (ISM) from which 
the solar nebula was formed had taken place $\sim 10^8$~yr earlier (cf.
Schramm \& Wasserburg 1970). 
This result can be obtained by considering two extreme cases
for the $r$-process production history prior to SSF:
(1) continuous uniform production 
(CUP, i.e., with an infinitesimal interval between successive events) with
the last event occurring $(\delta t)_{\rm CUP}$ years before SSF, and
(2) a single production (SP) event occurring
$(\delta t)_{\rm SP}$ years before SSF. In the CUP case, we have
\begin{equation}
\left({^{129}{\rm I}\over{^{127}{\rm I}}}\right)_{\rm SSF}=
\left({Y_{129}\over Y_{127}}\right)
\left({\bar\tau_{129}\over T_{\rm UP}}\right)
\exp[-(\delta t)_{\rm CUP}/\bar\tau_{129}],
\label{cup}
\end{equation}
where $Y_A$ represents the yield of the nuclide ``$A$'' in a single event
and $T_{\rm UP}$ is the period of uniform production.
For $Y_{129}/Y_{127}=1$ and $T_{\rm UP}=10^{10}$~yr, equation (\ref{cup})
gives $(\delta t)_{\rm CUP}=7.09\times 10^7$~yr.
In the SP case, we have
\begin{equation}
\left({^{129}{\rm I}\over{^{127}{\rm I}}}\right)_{\rm SSF}=
\left({Y_{129}\over Y_{127}}\right)
\exp[-(\delta t)_{\rm SP}/\bar\tau_{129}],
\label{sp}
\end{equation}
which gives $(\delta t)_{\rm SP}=2.09\times 10^8$~yr for $Y_{129}/Y_{127}=1$.
Were there only a single type of $r$-process sources for both $^{129}$I and
$^{182}$Hf, the abundance ratio $(^{182}{\rm Hf}/{^{180}{\rm Hf}})_{\rm SSF}$
(with the use of the solar $r$-process abundance of $^{182}$W 
and the solar abundance of $^{180}$Hf given in Table 1)
would have decayed to values of $4.27\times 10^{-8}$ (SP) to 
$2.28\times 10^{-6}$ (CUP) on the timescales deduced for the last injection
of $^{129}$I. Such values are
in clear conflict with the recently observed value of
$(^{182}{\rm Hf}/{^{180}{\rm Hf}})_{\rm SSF}=2.4\times 10^{-4}$ in meteorites
(Harper \& Jacobsen 1996; Lee \& Halliday 1995, 1996, 1997, 1999). Therefore,
the meteoritic data on $^{129}$I and $^{182}$Hf require at least
two distinct types of $r$-process events.

As the solar system abundances 
represent an average of Galactic chemical evolution on larger scales, 
it is reasonable to generalize the implications of
the meteoritic data discussed above 
to the larger-scale problem of the ``$r$-process.'' Using this 
approach, Wasserburg et al. (WBG96) 
concluded that there had to be at least two types of 
$r$-process events, one (H) occurring at a high frequency and one (L)
at a low frequency. The recurrence timescale for the
high frequency events was shown to be $\sim 10^7$ yr,
commensurate with the timescale for replenishment of a typical 
molecular cloud with fresh Type II supernova 
(hereafter referred to simply as supernova) debris.
They further pointed out that there should be
distinct differences in the $r$-process abundance peaks in very metal-poor
stars relative to the solar $r$-pattern. 
In particular, they inferred that there would be a dominance 
of the hypothesized high frequency supernova 
sources at early times with an abundance excess of the Pt peak at $A\sim 195$
relative to the $A\sim 130$ peak below Ba. 

Qian, Vogel \& Wasserburg (1998, hereafter QVW98) have
shown that in a two-component model to account for the solar $r$-pattern
and accommodate the meteoritic data on $^{129}$I and $^{182}$Hf at the same
time, the nuclei between the peak at $A\sim 130$ and the one at $A\sim 195$
are always produced along with the latter peak. In addition, it is not 
readily possible to produce the $A\sim 130$ peak without substantially
populating the region beyond this peak, especially when $\nu_e$ capture
on nuclei plays a significant role. They also found that
the total mass yield of the low frequency supernova L sources responsible
for the low mass peak at $A\sim 130$
must be $\sim 10$ times that of the high frequency H sources mainly
producing high mass nuclei beyond $A\sim 130$. 
They further speculated that the difference between the H and L sources 
was associated with the frequent formation of a black hole
in supernova H events (resulting in $\sim 5\times 10^8$ black 
holes with masses $\sim 1\,M_\odot$ in the present Galaxy)
and the less common production of a neutron star in the L events 
(resulting in a present Galactic inventory of $\sim 5\times 10^7$ 
neutron stars).

As mentioned earlier, stellar abundances of elements associated with
both the $A\sim 130$ and 195 $r$-process peaks at very low metallicities
are under active investigation (e.g., Sneden et al. 1996, 1998, 1999).
A preliminary report made by Cowan \& Sneden (1999) suggests that
more than one type of $r$-process events may be required.
For comparison with stellar observations,
it is particularly useful to present specific quantitative
inferences for the $r$-process elemental abundances
in very metal-poor stars based on the meteoritic data discussed above.
It will be shown that a two-component model has specific predictions 
for both the relative abundance patterns and the absolute abundances 
of $r$-process elements in stars formed 
from an ISM seeded with the ejecta from an individual 
supernova. A criterion for such earliest formed stars will be given based 
on the Eu abundance. It will further be argued that the $r$-process
abundances in many very metal-poor
stars are the result of many supernova contributions 
and that the ``metallicity'' 
[Fe/H] cannot provide a means of identifying the 
earliest formed stars.

This paper is organized as follows. In \S2, we show how the characteristics
of a two-component $r$-process model, such as the fractions of
$^{127}$I and $^{182}$W contributed by
the H and L events to the total solar $r$-process abundances of
these two nuclei, are determined by the abundance ratios
$(^{129}{\rm I}/{^{127}{\rm I}})_{\rm SSF}$ and 
$(^{182}{\rm Hf}/{^{180}{\rm Hf}})_{\rm SSF}$ in
meteorites. Two scenarios for the uniform production of $r$-process
nuclei relevant for the solar system abundances are discussed to provide
bounds on the proposed model. In \S3, we determine the yields in the H and
L events in our model from the solar $r$-process abundances by choosing
the yield templates in the low and high mass regions based on the observed
$r$-pattern in very metal-poor stars. We compare these predicted yields 
with the observed stellar abundances at very low metallicities in \S4 and
give our conclusions in \S5. Appendix A contains more general discussion
of a two-component $r$-process model, which emphasizes again the 
well-defined results from this model.

\section{Characteristics of a two-component $r$-process model}

In this paper, we will only consider addition of supernova debris 
to a ``standard'' mass of the ISM (see WBG96 and QVW98) 
without discussing the change in abundances due to astration 
that will store matter in stars over different timescales. 
We assume that the solar $r$-pattern is the result of two distinct types of
supernovae that occur at different frequencies over 
a time $T_{\rm UP}$ in the Galactic history preceding solar 
system formation (SSF). With the relative supernova yield of
$^{235}$U to $^{238}$U approximated by the 
ratio of the number of precursors for these two nuclei
(cf. Fowler \& Hoyle 1960; Fowler 1961), the estimate of the abundance ratio
$(^{235}{\rm U}/{^{238}{\rm U}})_{\rm SSF}$
based on a uniform rate of nucleosynthesis agrees very well with
the observed value of $(^{235}{\rm U}/{^{238}{\rm U}})_{\rm SSF}=0.317$
(e.g., Qian, Vogel, \& Wasserburg 1999).
This applies for uniform production timescales 
$T_{\rm UP}\approx 10^{10}$~yr, 
which are longer than the lifetime of $^{238}$U 
($\bar\tau_{238}=6.45\times 10^9$~yr) and much longer than that of $^{235}$U
($\bar\tau_{235}=1.02\times 10^9$~yr).
If the actinide production rate is estimated to be of the form 
$\exp(-t/\bar\tau_p)$, we find that the rate changes only by
a factor of $\sim 2$ over $10^{10}$~yr
(i.e., $\bar\tau_p\sim 1.3\times 10^{10}$~yr)
in order to account for the observed
value of $(^{235}{\rm U}/{^{238}{\rm U}})_{\rm SSF}$.
We thus adopt a model of uniform production (cf. WBG96; Qian et al. 1999). 

There are three issues involved in considering the early solar
system abundance of short-lived nuclei derived from supernovae. These are
the yield in a supernova, the dilution factor of the ISM, and the time 
difference between 
the last supernova event and the formation of the solar system. 
We will discuss two scenarios, in which the last supernova event
is either somewhat earlier than (scenario A) or coincident with 
(scenario B, i.e., a trigger for)
the formation of the solar system.
These two scenarios are considered as bounds to the proposed model.

\subsection{Scenario A}

For simplicity, let us 
choose a standard dilution factor for all supernovae
(cf. Tsujimoto \& Shigeyama 1998; Qian et al. 1999) and
focus on a specific volume of the ISM
corresponding to the standard mass diluting the debris 
from an individual supernova. 
Over a time $T_{\rm UP}$ in the Galactic history preceding SSF,
supernovae in this ``volume'' can inject fresh nucleosynthesis products
to give the abundances of radioactive nuclei in the early solar system and
those of stable nuclei in the present solar system (cf. WBG96; QVW98). 
Let us assume that
a supernova $r$-process event regularly occurs every 
$\Delta$ years in this volume with the last
event taking place $\Delta$ years prior to SSF (scenario A). 
In this scenario, SSF occurs just before a new injection of $r$-process
nuclei into the ISM and
$\Delta$ years after the most recent injection. 
Then for a short-lived nuclide ``${\cal{R}}$'' 
with $\bar\tau_{\cal{R}}\ll T_{\rm UP}$, 
the net number of ${\cal{R}}$ nuclei present at the time of SSF is
\begin{equation}
N_{\cal{R}}(t_{\rm SSF}) = \sum_{j=1}^{j_{\rm max}}Y_{\cal{R}} 
\exp(-\,j\,\Delta/\bar\tau_{\cal{R}})\approx 
Y_{\cal{R}}/[\exp(\Delta/\bar\tau_{\cal{R}})-1],
\end{equation}
where $Y_{\cal{R}}$ 
is the number of ${\cal{R}}$ nuclei produced 
per event (i.e., the yield) and is assumed to be constant. 
Note that for $\Delta\ll \bar\tau_{\cal{R}}$, we have
\begin{equation}
N_{\cal{R}}(t_{\rm SSF})\approx 
Y_{\cal{R}}{\bar\tau_{\cal{R}}\over\Delta}
= Y_{\cal{R}}\bar f\bar\tau_{\cal{R}},
\end{equation}
where $\bar f=1/\Delta$ is the frequency. However,
for $\Delta\gg \bar\tau_{\cal{R}}$, we have 
$N_{\cal{R}}(t_{\rm SSF})\approx Y_{\cal{R}} 
\exp(-\Delta/\bar\tau_{\cal{R}})$ and the last event dominates at SSF. 
Obviously, for a stable nuclide ``${\cal{S}}$,'' 
the number of ${\cal{S}}$ nuclei present
at the time of SSF is
\begin{equation}
N_{\cal{S}}(t_{\rm SSF})=Y_{\cal{S}}{T_{\rm UP}\over\Delta},
\end{equation}
where $Y_{\cal{S}}$ is the number of ${\cal{S}}$ nuclei produced 
per event and should be about 
the same as the yield of the corresponding short-lived nuclide.
We now turn to the case of two distinct $r$-process components.

\subsubsection{The limiting case}

First consider the limiting case where the L events occurring at
a frequency $\bar f_{\rm L}=1/\Delta_{\rm L}$ 
produce all of the $^{129}$I and $^{127}$I but no $^{182}$Hf,
while the H events occurring at a frequency 
$\bar f_{\rm H}=1/\Delta_{\rm H}$ 
produce all of the $^{182}$Hf (hence
all of the $r$-process contribution to its stable daughter $^{182}$W) 
but no $^{129}$I or $^{127}$I. The meaning of
L and H will become clear shortly. In this limiting case, we have
\begin{equation}
\left({^{129}{\rm I}\over{^{127}{\rm I}}}\right)_{\rm SSF}=
\left({Y_{129}\over Y_{127}}\right)
\left({\bar\tau_{129}\over T_{\rm UP}}\right)
{\Delta_{\rm L}/\bar\tau_{129}\over\exp(\Delta_{\rm L}/\bar\tau_{129})-1},
\label{l0}
\end{equation}
and
\begin{equation}
\left({^{182}{\rm Hf}\over{^{182}{\rm W}_r}}\right)_{\rm SSF}=
\left({\bar\tau_{182}\over T_{\rm UP}}\right)
{\Delta_{\rm H}/\bar\tau_{182}\over\exp(\Delta_{\rm H}/\bar\tau_{182})-1},
\label{h0}
\end{equation}
where $^{182}{\rm W}_r$ stands for the $r$-process contribution to the
solar inventory of $^{182}$W.
Equations (\ref{l0}) and (\ref{h0}) can be reduced to the form
\begin{equation}
f(X_{129}^{\rm L})=C_{\rm I},
\label{l1}
\end{equation}
and
\begin{equation}
f(X_{182}^{\rm H})=C_{\rm Hf},
\label{h1}
\end{equation}
when we define $X_{129}^{\rm L} \equiv \Delta_{\rm L}/\bar\tau_{129}$,
$X_{182}^{\rm H} \equiv \Delta_{\rm H}/\bar\tau_{182}$,
\begin{equation}
f(X)\equiv{X\over\exp(X)-1},
\end{equation}
\begin{equation}
C_{\rm I}\equiv{(^{129}{\rm I}/{^{127}{\rm I}})_{\rm SSF}\over
(Y_{129}/Y_{127})(\bar\tau_{129}/T_{\rm UP})},
\label{ci}
\end{equation}
and
\begin{equation}
C_{\rm Hf}\equiv{(^{182}{\rm Hf}/{^{182}{\rm W}_r})_{\rm SSF}\over
\bar\tau_{182}/T_{\rm UP}}.
\label{chf}
\end{equation}

Both $^{129}$I and $^{127}$I are essentially pure $r$-process nuclei.
As these two nuclei are so close in mass number, the relative yield
$Y_{129}/Y_{127}$ in the L events must be close to unity.
For clarity of presentation, we choose $Y_{129}/Y_{127}=1$ in the following
discussion. However,
we have checked that all of the results obtained in this paper are 
insensitive to variations of $Y_{129}/Y_{127}$ between 1 and 2.
On the scale of $N_\odot({\rm Si})=10^6$, the solar abundance of $^{180}$Hf
is $N_\odot(^{180}{\rm Hf})=0.0541$ (Anders \& Grevesse 1989) and the solar
$r$-process abundance of $^{182}$W is $N_{\odot,r}(^{182}{\rm W})=0.0222$
(K\"appeler et al. 1991; Arlandini et al. 1999). This gives 
$(^{182}{\rm Hf}/{^{182}{\rm W}_r})_{\rm SSF}=5.85\times 10^{-4}$.
With the above information, we obtain $C_{\rm I}=0.0441$ and
$C_{\rm Hf}=0.450$ for $T_{\rm UP}=10^{10}$~yr 
(assumed throughout this paper). These and other data pertinent to our 
discussion are summarized in Table 1. 

For the numerical values of $C_{\rm I}$ and $C_{\rm Hf}$ chosen above,
the solutions to equations (\ref{l1}) and (\ref{h1}) give
$\Delta_{\rm L}=1.06\times 10^8$~yr and $\Delta_{\rm H}=1.85\times 10^7$~yr.
Thus, L can be understood to stand for low frequency and H for high frequency
$r$-process events.

\subsubsection{The general case}

In general, a two-component model does not require that the H events 
produce none of the low mass ($A\lesssim 130$) nuclei associated with
$^{127}$I or that the L events
produce no high mass ($A>130$) nuclei associated with $^{182}$W. 
It only requires
that the relative yields in these two mass regions be different (cf. QVW98). 
For the general two-component model, equation (\ref{l0}) is replaced by
\begin{equation}
\left({^{129}{\rm I}\over{^{127}{\rm I}}}\right)_{\rm SSF}=
\left({Y_{129}^{\rm L}\over Y_{127}^{\rm L}}\right)
\left({\bar\tau_{129}\over T_{\rm UP}}\right)
{f(X_{129}^{\rm H})+(Y_{129}^{\rm L}/Y_{129}^{\rm H})
(\Delta_{\rm H}/\Delta_{\rm L})f(X_{129}^{\rm L})
\over (Y_{127}^{\rm H}/Y_{127}^{\rm L})(Y_{129}^{\rm L}/Y_{129}^{\rm H})
+(Y_{129}^{\rm L}/Y_{129}^{\rm H})(\Delta_{\rm H}/\Delta_{\rm L})},
\label{la0}
\end{equation}
where we have defined $X_{129}^{\rm H}\equiv\Delta_{\rm H}/\bar\tau_{129}$ and
$X_{129}^{\rm L}\equiv\Delta_{\rm L}/\bar\tau_{129}$. 
If we assume $Y_{129}^{\rm L}/Y_{127}^{\rm L}=
Y_{129}^{\rm H}/Y_{127}^{\rm H}=Y_{129}/Y_{127}$, then
equation (\ref{la0}) becomes 
\begin{equation}
{f(X_{129}^{\rm H})+(Y_{129}^{\rm L}/Y_{129}^{\rm H})
(\Delta_{\rm H}/\Delta_{\rm L})f(X_{129}^{\rm L})\over
1+(Y_{129}^{\rm L}/Y_{129}^{\rm H})(\Delta_{\rm H}/\Delta_{\rm L})}
=C_{\rm I},
\label{la1}
\end{equation}
which can be rewritten as
\begin{equation}
\left({Y_{129}^{\rm L}\over Y_{129}^{\rm H}}\right)
\left({\Delta_{\rm H}\over\Delta_{\rm L}}\right)=
\left({Y_{127}^{\rm L}\over Y_{127}^{\rm H}}\right)
\left({\Delta_{\rm H}\over\Delta_{\rm L}}\right)=
{f(X_{129}^{\rm H})-C_{\rm I}\over C_{\rm I}-f(X_{129}^{\rm L})}.
\label{la2}
\end{equation}
Similarly, we obtain
\begin{equation}
\left({Y_{182}^{\rm L}\over Y_{182}^{\rm H}}\right)
\left({\Delta_{\rm H}\over\Delta_{\rm L}}\right)=
{f(X_{182}^{\rm H})-C_{\rm Hf}\over C_{\rm Hf}-f(X_{182}^{\rm L})},
\label{ha2}
\end{equation}
where we have defined $X_{182}^{\rm H}\equiv\Delta_{\rm H}/\bar\tau_{182}$ and
$X_{182}^{\rm L} \equiv \Delta_{\rm L}/\bar\tau_{182}$.

By definition, we have $\Delta_{\rm H}<\Delta_{\rm L}$. Physical solutions
(i.e., positive values for the yield ratios
$Y_{127}^{\rm L}/Y_{127}^{\rm H}$ and $Y_{182}^{\rm L}/Y_{182}^{\rm H}$) 
to equations (\ref{la2}) and
(\ref{ha2}) then require 
$\Delta_{\rm L}> 1.06\times 10^8~{\rm yr}\equiv\Delta_{\rm L}^{\rm min}$ and
$\Delta_{\rm H}< 1.85\times 10^7~{\rm yr}\equiv\Delta_{\rm H}^{\rm max}$
(hence $\Delta_{\rm H}/\Delta_{\rm L}< 0.175$). Note that
$\Delta_{\rm L}^{\rm min}$ and $\Delta_{\rm H}^{\rm max}$ correspond to
the values of $\Delta_{\rm L}$ and $\Delta_{\rm H}$ in the limiting case.
The interrelationship between $\Delta_{\rm H}$ and $\Delta_{\rm L}$  
and the yield ratios $Y_{127}^{\rm L}/Y_{127}^{\rm H}$ and
$Y_{182}^{\rm L}/Y_{182}^{\rm H}$ can be understood as follows. 
The production of $^{129}$I
and $^{127}$I in the H events
mainly increases the inventory of 
the radioactive $^{129}$I at SSF
as the last few H events occurred much closer to the 
formation of the early solar system than the last L event.
In order to satisfy the observed value of 
$(^{129}{\rm I}/{^{127}{\rm I}})_{\rm SSF}$,
the last L event must then move to even earlier times, i.e., 
$\Delta_{\rm L}$ must increase from that in the limiting case.
On the other hand, the production of any 
$^{182}$Hf (hence $^{182}$W) in the L events 
mainly increase the stable
inventory of $^{182}{\rm W}_r$ at SSF. In order to satisfy 
the abundance ratio
$(^{182}{\rm Hf}/{^{182}{\rm W}_r})_{\rm SSF}$, 
this then requires a decrease in
$\Delta_{\rm H}$ from that in the limiting case to allow more contribution
to the radioactive $^{182}$Hf from the last few H events 
to compensate for the increased portion of the $^{182}{\rm W}_r$ 
inventory present from the L events.

In addition to equations (\ref{la2}) and (\ref{ha2}), a further requirement
of the two-component model is that the integrated $r$-process
production of the stable ${\cal{S}}$
nuclei must satisfy
\begin{equation}
{N_{127}^{\rm H}(t_{\rm SSF})+N_{127}^{\rm L}(t_{\rm SSF})\over
N_{182}^{\rm H}(t_{\rm SSF})+N_{182}^{\rm L}(t_{\rm SSF})}=
\left({Y_{127}^{\rm H}\over Y_{182}^{\rm H}}\right)
{1+(Y_{127}^{\rm L}/Y_{127}^{\rm H})(\Delta_{\rm H}/\Delta_{\rm L})\over
1+(Y_{182}^{\rm L}/Y_{182}^{\rm H})(\Delta_{\rm H}/\Delta_{\rm L})}=
{N_{\odot,r}(^{127}{\rm I})\over N_{\odot,r}(^{182}{\rm W})},
\label{sa}
\end{equation}
where $N_{\odot,r}(^{127}{\rm I})\approx N_\odot(^{127}{\rm I})$ and
$N_{\odot,r}(^{182}{\rm W})$ are the solar $r$-process abundances of
$^{127}$I and $^{182}$W [given in Table 1 on the scale of
$N_\odot({\rm Si})=10^6$], respectively.

The characteristics of scenario A are fully specified by equations
(\ref{la2})--(\ref{sa}). For convenience in discussing 
the solutions to these equations, we define
\begin{equation}
F_r^{\rm L}(^{127}{\rm I})\equiv
{N_{127}^{\rm L}(t_{\rm SSF})\over N_{127}^{\rm H}(t_{\rm SSF})+
N_{127}^{\rm L}(t_{\rm SSF})}=1-
{1\over 1+(Y_{127}^{\rm L}/Y_{127}^{\rm H})(\Delta_{\rm H}/\Delta_{\rm L})},
\label{fli}
\end{equation} 
which is the fraction of $^{127}$I contributed by the L events to 
the corresponding total solar $r$-process
abundance, and
\begin{equation}       
F_r^{\rm H}(^{182}{\rm W})\equiv
{N_{182}^{\rm H}(t_{\rm SSF})\over N_{182}^{\rm H}(t_{\rm SSF})+
N_{182}^{\rm L}(t_{\rm SSF})}=	
{1\over 1+(Y_{182}^{\rm L}/Y_{182}^{\rm H})(\Delta_{\rm H}/\Delta_{\rm L})},
\label{fhw}
\end{equation}
which is the fraction of $^{182}$W contributed by the H events to the 
corresponding total 
solar $r$-process abundance. 
The limiting cases outlined in \S2.1.1 correspond to
$F_r^{\rm L}(^{127}{\rm I})=1$, for which
no $^{129}$I or $^{127}$I is produced in the H events, 
and $F_r^{\rm H}(^{182}{\rm W})=1$, for which
no $^{182}$Hf (hence no $^{182}$W) is produced in the L events. 
We also define
\begin{equation}
U_{\rm H}\equiv{Y_{127}^{\rm H}/Y_{182}^{\rm H}\over
N_{\odot,r}(^{127}{\rm I})/N_{\odot,r}(^{182}{\rm W})},
\label{uh}
\end{equation}
which measures the relative yield of $^{127}$I to
$^{182}$W in an H event
with respect to the corresponding solar $r$-process abundance ratio. 
When $U_{\rm H}< 1$, the production of 
$^{127}$I relative to $^{182}$W in an H event is depleted (subsolar)
compared with the solar $r$-pattern. 
Note that similarly defined quantities can be obtained as
\begin{equation}
F_r^{\rm H}(^{127}{\rm I}))\equiv
{N_{127}^{\rm H}(t_{\rm SSF})\over N_{127}^{\rm H}(t_{\rm SSF})+
N_{127}^{\rm L}(t_{\rm SSF})}
=1-F_r^{\rm L}(^{127}{\rm I}),
\label{fhi}
\end{equation}
\begin{equation}
F_r^{\rm L}(^{182}{\rm W})\equiv
{N_{182}^{\rm L}(t_{\rm SSF})\over N_{182}^{\rm H}(t_{\rm SSF})+
N_{182}^{\rm L}(t_{\rm SSF})}
=1-F_r^{\rm H}(^{182}{\rm W}),
\label{flw}
\end{equation}
and
\begin{equation}
U_{\rm L}\equiv{Y_{127}^{\rm L}/Y_{182}^{\rm L}\over
N_{\odot,r}(^{127}{\rm I})/N_{\odot,r}(^{182}{\rm W})}
= U_{\rm H}{(Y_{127}^{\rm L}/Y_{127}^{\rm H})\over
(Y_{182}^{\rm L}/Y_{182}^{\rm H})}.
\label{ul}
\end{equation}

A unique solution for $\Delta_{\rm H}$, $\Delta_{\rm L}$,
$Y_{127}^{\rm L}/Y_{127}^{\rm H}$, $Y_{182}^{\rm L}/Y_{182}^{\rm H}$, and
$Y_{127}^{\rm H}/Y_{182}^{\rm H}$
is not possible to obtain from equations (\ref{la2})--(\ref{sa}). 
However, substantial restrictions on all of these parameters that
exhibit the basic characteristics of scenario A of the two-component model
can be obtained by only considering the constraints 
$\Delta_{\rm H}< \Delta_{\rm H}^{\rm max}=1.85\times 10^7$~yr and
$\Delta_{\rm L}> \Delta_{\rm L}^{\rm min}=1.06\times 10^8$~yr on
equations (\ref{la2}) and (\ref{ha2}) (see Appendix A). We will now 
use a clear example for our discussion of this scenario. We assume that
$\Delta_{\rm H}/\Delta_{\rm L} = 0.1$, which specifies a physical range of
(1.06 to $1.85)\times 10^8$~yr for $\Delta_{\rm L}$.
The key parameters $U_{\rm H}$, $U_{\rm L}$,
$F_r^{\rm L}(^{127}{\rm I})$, and
$F_r^{\rm H}(^{182}{\rm W})$ obtained for this example
are shown in Figure 1 as functions of $\Delta_{\rm L}$.
The relative yield of $^{127}$I to $^{182}$W in the L events is 3.28 times
the corresponding solar $r$-process abundance ratio ($U_{\rm L}= 3.28$, i.e.,
supersolar) at $\Delta_{\rm L}=\Delta_{\rm L}^{\rm min}=1.06\times 10^8$~yr
and increases as the limiting case $\Delta_{\rm H}=\Delta_{\rm H}^{\rm max}$
is approached. By contrast, 
the relative yield of $^{127}$I to $^{182}$W in the 
H events is always more than a factor of 15 less than the corresponding solar 
$r$-process abundance ratio ($U_{\rm H}\leq 0.065$, i.e., subsolar).
For $\Delta_{\rm L}=1.85\times 10^8$~yr corresponding to
$\Delta_{\rm H}=\Delta_{\rm H}^{\rm max}$, 
93.5\% of the $^{127}$I is produced 
by the L events [$F_r^{\rm L}(^{127}{\rm I})=0.935$] and all of the solar
$r$-process $^{182}$W is produced by the H events 
[$F_r^{\rm H}(^{182}{\rm W})=1$]. 
For $\Delta_{\rm L}=\Delta_{\rm L}^{\rm min}$, all of the $^{127}$I 
is produced by the L events [$F_r^{\rm L}(^{127}{\rm I})=1$] and
69.5\% of the solar $r$-process $^{182}$W is produced by the H events
[$F_r^{\rm H}(^{182}{\rm W})=0.695$]. Note that in the above example for
scenario A, the fraction of $^{127}$I
contributed by the H events to the corresponding
total solar $r$-process abundance 
is at most 6.5\% [$F_r^{\rm H}(^{127}{\rm I})\leq 0.065$].
On the other hand, the fraction of $^{182}$W
contributed by the L events to the corresponding total
solar $r$-process abundance is 
$F_r^{\rm L}(^{182}{\rm W})\geq 0.1$ 
for $\Delta_{\rm L}< 1.63\times 10^8$~yr and can be as much as 30.5\% at
$\Delta_{\rm L}=\Delta_{\rm L}^{\rm min}$. 
This means that for most of the physical
$\Delta_{\rm L}$ values corresponding to 
$\Delta_{\rm H}/\Delta_{\rm L}=0.1$ in scenario A, the yields of $^{182}$W 
in the H and L events are comparable, i.e., 
$Y_{182}^{\rm L}/Y_{182}^{\rm H}\sim 1$.

In deriving the above results on $^{127}$I, we have assumed that
$Y_{129}^{\rm H}/Y_{127}^{\rm H}=Y_{129}^{\rm L}/Y_{127}^{\rm L}$
(cf. eqs. [\ref{la0}] and [\ref{la1}]).
We note that $Y_{129}^{\rm H}/Y_{127}^{\rm H}$ may differ from
$Y_{129}^{\rm L}/Y_{127}^{\rm L}$. For example, the yields for
$A\lesssim 130$ in the H events may drop steeply at smaller $A$
(cf. QVW98). Consider the extreme case where $Y_{127}^{\rm H}=0$, for which
the denominator on the right-hand side of equation (\ref{la0}) would be
$(Y_{129}^{\rm L}/Y_{129}^{\rm H})(\Delta_{\rm H}/\Delta_{\rm L})$.
The correct equation to be used in this case would be identical to
equation (\ref{la1}) except for the unity term in the denominator. 
However, it is clear that
practically all of the low mass nuclei are produced in the L events, i.e.,
$(Y_{129}^{\rm L}/Y_{129}^{\rm H})(\Delta_{\rm H}/\Delta_{\rm L})\gg 1$.
Therefore, our assumption of 
$Y_{129}^{\rm H}/Y_{127}^{\rm H}=Y_{129}^{\rm L}/Y_{127}^{\rm L}$
has little effect on estimating the production
of low mass nuclei in the L events. On the other hand, the
ratio $Y_{127}^{\rm L}/Y_{127}^{\rm H}$ derived from equation (\ref{la2})
may be regarded as a safe lower
limit for $Y_A^{\rm L}/Y_A^{\rm H}$ at $A<130$, although we will later use it
to represent the actual yield ratios in the low mass region.

\subsection{Scenario B}

We now discuss the second scenario which differs from scenario A
in that the solar system is assumed to have been
formed {\it immediately} after
the last H event (scenario B). The standard dilution factor of the ISM is
assumed for the debris from this last event.
At the time of SSF in scenario B, the net number of short-lived
${\cal{R}}$ nuclei due to the H events is
\begin{equation}
N_{\cal{R}}^{\rm H}(t_{\rm SSF}) = 
\sum_{j=0}^{j_{\rm max}}Y_{\cal{R}}^{\rm H}
\exp(-\,j\,\Delta/\bar\tau_{\cal{R}})\approx
Y_{\cal{R}}^{\rm H}/[1-\exp(-\Delta/\bar\tau_{\cal{R}})].
\end{equation}
Accordingly, we define
\begin{equation}
g(X)\equiv{X\over 1-\exp(-X)},
\end{equation}
and replace equations (\ref{la2}) and (\ref{ha2}) by
\begin{equation}
\left({Y_{127}^{\rm L}\over Y_{127}^{\rm H}}\right)
\left({\Delta_{\rm H}\over\Delta_{\rm L}}\right)=
{g(X_{129}^{\rm H})-C_{\rm I}\over C_{\rm I}-f(X_{129}^{\rm L})},
\label{lb2}
\end{equation}
and
\begin{equation}
\left({Y_{182}^{\rm L}\over Y_{182}^{\rm H}}\right)
\left({\Delta_{\rm H}\over\Delta_{\rm L}}\right)=
{g(X_{182}^{\rm H})-C_{\rm Hf}\over C_{\rm Hf}-f(X_{182}^{\rm L})},
\label{hb2}
\end{equation}
respectively. The characteristics of scenario B are then fully specified by 
equations (\ref{lb2}), (\ref{hb2}), and (\ref{sa}).

As $g(X)\geq 1$ for $X\geq 0$, physical solutions to equations (\ref{lb2})
and (\ref{hb2}) only require that 
$\Delta_{\rm L} > \Delta_{\rm L}^{\rm min}=1.06\times 10^8$~yr.
Unlike in scenario A, there is no restriction on $\Delta_{\rm H}$ in
scenario B. However, some general characteristics of this scenario
can still be obtained without any detailed knowledge of $\Delta_{\rm H}$.
As $f(X_{129}^{\rm L}) > 0$, equation (\ref{lb2}) gives
\begin{equation}
1+\left({Y_{127}^{\rm L}\over Y_{127}^{\rm H}}\right)
\left({\Delta_{\rm H}\over\Delta_{\rm L}}\right) >
{g(X_{129}^{\rm H})\over C_{\rm I}}\geq {1\over  C_{\rm I}},
\label{lb3}
\end{equation}
which means that $F_r^{\rm H}(^{127}{\rm I}) < C_{\rm I}=0.044$ and
$F_r^{\rm L}(^{127}{\rm I}) > 0.956$. Thus the H events provide 
less than 4.4\% of the solar $r$-process $^{127}$I in scenario B.

For $\Delta_{\rm L} > \Delta_{\rm L}^{\rm min}=1.06\times 10^8$~yr,
we have $f(X_{182}^{\rm L}) < 2.35\times 10^{-3}\ll C_{\rm Hf}$.
Equation (\ref{hb2}) then gives
\begin{equation}
1+\left({Y_{182}^{\rm L}\over Y_{182}^{\rm H}}\right)
\left({\Delta_{\rm H}\over\Delta_{\rm L}}\right)\approx
{g(X_{182}^{\rm H})\over C_{\rm Hf}}\geq {1\over C_{\rm Hf}},
\label{hb3}
\end{equation}
which means that $F_r^{\rm H}(^{182}{\rm W})\lesssim C_{\rm Hf}=0.45$ and
$F_r^{\rm L}(^{182}{\rm W})\gtrsim 0.55$. 
Therefore, the L events account for
essentially all of the solar $r$-process
$^{127}$I and more than 55\% of the solar $r$-process
$^{182}$W in scenario B.

To give a specific example for scenario B, we again choose
$\Delta_{\rm H}/\Delta_{\rm L}=0.1$ and further impose
$\Delta_{\rm L}\leq 2\times 10^8$~yr [near the $(\delta t)_{\rm SP}$ value
for the single production case in eq. (\ref{sp})]. 
The key parameters $U_{\rm H}$,
$U_{\rm L}$, $F_r^{\rm L}(^{127}{\rm I})$, and $F_r^{\rm L}(^{182}{\rm W})$
obtained for this example
are shown as functions of $\Delta_{\rm L}$ in Figure 2.
In accord with the above general discussion of scenario B,
the L events in this example 
provide (100 to 97)\% of the solar $r$-process $^{127}$I
[$F_r^{\rm L}(^{127}{\rm I})=1$ to 0.97] and (69 to 77)\% of
the solar $r$-process $^{182}$W
[$F_r^{\rm L}(^{182}{\rm W})=0.69$ to 0.77] over the range of
(1.06 to $2)\times 10^8$~yr for $\Delta_{\rm L}$.
As a result, the relative yield of $^{127}$I to $^{182}$W 
in an L event in scenario B is
very close to the corresponding
solar $r$-process abundance ratio
($U_{\rm L}=1.44$ to 1.26 in Fig. 2).

\section{Yields in a two-component $r$-process model}

Let us now consider the yields in individual H and L events
in a two-component $r$-process model.
For a stable nuclide ${\cal{S}}$, the total $r$-process abundance of 
${\cal{S}}$ nuclei at the time of SSF is
\begin{equation}
N_{\cal{S}}(t_{\rm SSF})=Y_{\cal{S}}^{\rm H}{T_{\rm UP}\over\Delta_{\rm H}}
+ Y_{\cal{S}}^{\rm L}{T_{\rm UP}\over\Delta_{\rm L}} 
= {Y_{\cal{S}}^{\rm H}\over F_r^{\rm H}({\cal{S}})}
\left({T_{\rm UP}\over\Delta_{\rm H}}\right),
\label{ys}
\end{equation}
where as in \S2, the parameter 
\begin{equation}
F_r^{\rm H}({\cal{S}})={1\over 1+ (Y_{\cal{S}}^{\rm L}/Y_{\cal{S}}^{\rm H})
(\Delta_{\rm H}/\Delta_{\rm L})}
\label{fhs}
\end{equation}
is the fraction of the nuclide ${\cal{S}}$ contributed by the H events 
to the total solar $r$-process
abundance of ${\cal{S}}$ nuclei.
Assuming that the diluting number of hydrogen (${\cal{H}}$) atoms 
per event is a constant $N_{\cal{H}}$, we have
$N_{\cal{S}}(t_{\rm SSF})/N_{\cal{H}}
=N_{\odot,r}({\cal{S}})/N_\odot({\cal{H}})$ and
\begin{equation}
\log\left({Y_{\cal{S}}^{\rm H}\over N_{\cal{H}}}\right)=
\log\left[{N_{\odot,r}({\cal{S}})\over N_\odot({\cal{H}})}\right]
+\log[F_r^{\rm H}({\cal{S}})]
-\log\left({T_{\rm UP}\over\Delta_{\rm H}}\right),
\label{yh}
\end{equation}
where $N_{\odot,r}({\cal{S}})$ is the solar $r$-process abundance of
${\cal{S}}$ nuclei and $N_\odot({\cal{H}})$ 
is the solar abundance of hydrogen. For a star formed from an ISM
contaminated by a single H event,
its abundance of ${\cal{S}}$ nuclei (with respect to hydrogen) is given by
equation (\ref{yh}).
In standard spectroscopic notation,
$\log\epsilon({\cal{S}})=\log({\cal{S}}/{\cal{H}})+12$,
where ${\cal{S}}/{\cal{H}}$ is the abundance ratio of the nuclide
${\cal{S}}$ to hydrogen in the star. 
So equation (\ref{yh}) for the H event can be rewritten as
\begin{equation}
\log\epsilon_{\rm H}({\cal{S}})=\log\epsilon_{\odot,r}({\cal{S}})
+\log[F_r^{\rm H}({\cal{S}})]
-\log(T_{\rm UP}/\Delta_{\rm H}).
\label{eh}
\end{equation}
Similarly, for the L event, we obtain
\begin{equation}
\log\epsilon_{\rm L}({\cal{S}})=\log\epsilon_{\odot,r}({\cal{S}})
+\log[F_r^{\rm L}({\cal{S}})]
-\log(T_{\rm UP}/\Delta_{\rm L}),
\label{el}
\end{equation}
where $F_r^{\rm L}({\cal{S}})=1-F_r^{\rm H}({\cal{S}})$ is the fraction of
the nuclide ${\cal{S}}$
contributed by the L events to the total solar $r$-process abundance of
${\cal{S}}$ nuclei. 
Note that when put in terms of the $\log\epsilon$ values for
a star formed from an ISM contaminated by a single H or L event,
the yields of ${\cal{S}}$ nuclei as 
given in equations (\ref{eh}) and (\ref{el})
only depend on the basic results
$F_r^{\rm H}({\cal{S}})$ [or equivalently,
$F_r^{\rm L}({\cal{S}})$], $T_{\rm UP}/\Delta_{\rm H}$, 
and $T_{\rm UP}/\Delta_{\rm L}$ from a two-component $r$-process model.

\subsection{Yields in a simplified two-component model}

With equations (\ref{eh}) and (\ref{el}), we can now relate the yields
in a two-component $r$-process model, 
and hence the abundances to be observed
in stars formed from an ISM contaminated by a single H or L event,
to the solar $r$-process abundances. 
The application to $^{127}$I and $^{182}$W is straightforward as their
yields are a direct consequence of the model involving the radioactive
nuclei $^{129}$I and $^{182}$Hf discussed in \S2.
In Figure 3, we show
$\delta\log\epsilon_{\rm H}({\cal{S}})\equiv
\log\epsilon_{\rm H}({\cal{S}})-\log\epsilon_{\odot,r}({\cal{S}})$ and
$\delta\log\epsilon_{\rm L}({\cal{S}})\equiv
\log\epsilon_{\rm L}({\cal{S}})-\log\epsilon_{\odot,r}({\cal{S}})$
for $^{127}$I and $^{182}$W as functions of $\Delta_{\rm L}$. 
The corresponding parameters $F_r^{\rm H}({\cal{S}})$ 
and $F_r^{\rm L}({\cal{S}})$
shown in Figures 1(b) and 2(b) have been used to obtain 
the curves in Figure 3.
Note that $\delta\log\epsilon_{\rm H}(^{182}{\rm W})$
is dominated by the term
$-\log(T_{\rm UP}/\Delta_{\rm H})$ in equation (\ref{eh}), and
$\delta\log\epsilon_{\rm L}(^{127}{\rm I})$
is dominated by the term
$-\log(T_{\rm UP}/\Delta_{\rm L})$ in equation (\ref{el}). For
$\Delta_{\rm H}/\Delta_{\rm L}=0.1$
(assumed in Figures 1 and 2) and $\Delta_{\rm L}=1.5\times 10^8$~yr, 
the contributions from these two terms are
$-2.82$ and $-1.82$, respectively. These are the 
minimum values of $\delta\log\epsilon_{\rm H}(^{182}{\rm W})$
and $\delta\log\epsilon_{\rm L}(^{127}{\rm I})$ to be
expected for the assumed values of $\Delta_{\rm H}$ and
$\Delta_{\rm L}$ as $F_r^{\rm H}(^{182}{\rm W})$ and
$F_r^{\rm L}(^{127}{\rm I})$ cannot exceed unity 
(cf. eqs. [\ref{eh}] and [\ref{el}]).

If one wishes to establish the yields of stable nuclei other than
$^{127}$I and $^{182}$W, then it is 
necessary to know the ratios $Y_{\cal{S}}^{\rm L}/Y_{\cal{S}}^{\rm H}$ of
their yields in an L event to those in an H event (cf. eq. [\ref{ys}]).
The simplest approach is to assume 
that the H and L events produce the low mass nuclei associated with
$^{127}$I and the high mass nuclei associated with $^{182}$W in
different proportions, but with each mass region having exactly
the same yield template as the corresponding solar $r$-pattern. 
In this case, it is only necessary to 
define the two mass regions. The yield ratios
$Y_{\cal{S}}^{\rm L}/Y_{\cal{S}}^{\rm H}$ for all the stable nuclei in the
low or high mass region are then specified by
$Y_{127}^{\rm L}/Y_{127}^{\rm H}$ or
$Y_{182}^{\rm L}/Y_{182}^{\rm H}$, respectively.
Likewise, the values of $\delta\log\epsilon_{\rm H}({\cal{S}})$ and
$\delta\log\epsilon_{\rm L}({\cal{S}})$ for all the stable nuclei in the
low or high mass region (cf. eqs. [\ref{fhs}], [\ref{eh}], and [\ref{el}])
are specified by the corresponding values for
$^{127}$I or $^{182}$W (shown in Fig. 3), respectively.
 
We choose to assign nuclei with $A\leq 130$, including $^{127}$I, to
the low mass region and those with $A > 130$, including $^{182}$W, to the
high mass region. This choice is justified by both the observed stellar
abundances at very low metallicities and 
the nuclear physics of the $r$-process.
Observational studies by 
Sneden et al. (1996, 1998, 1999) have established that the $r$-process
elemental abundance pattern for Ba ($A\sim 135$) and higher atomic numbers
in very metal-poor stars agrees very well with the corresponding solar
$r$-pattern. In two cases, Th was also detected and after
correction for the age of the star, the observed abundance of Th relative to
those of the stable $r$-process nuclei at lower atomic numbers (e.g., Eu) 
is again consistent with the solar $r$-pattern (cf. Cowan et al. 1999).
These studies include stars with metallicities of [Fe/H]
$=-1.67$ down to $-3.1$ although the data sets are not complete 
over the entire range of atomic numbers corresponding to $A > 130$.
In addition, the extensive work by McWilliam et al. (1995) and 
McWilliam (1998) has shown that the abundance ratio of 
Ba to Eu ($A\sim 151$) at very low
metallicities is remarkably well-behaved and lies in a very narrow band
around the corresponding solar $r$-process value (see \S4 and Fig. 9).
Therefore, it is reasonable to conclude from these observational studies
that apart from some small 
discrepancies (e.g., at Ce, see Sneden et al. 1998 and Fig. 8), 
the yield pattern for $A > 130$ in all
$r$-process events is very close to the solar $r$-pattern 
in this mass region. 

Further, the above observational result can be understood
from the nuclear physics of the $r$-process. Consider the making of the Pt
peak at $A\sim 195$ by rapid neutron capture onto seed nuclei with
$A\lesssim 100$. During the $r$-process, the nuclear flow is mainly 
controlled by $\beta$-decay of the progenitor nuclei. The time needed for
the nuclear flow to clear a mass region is approximately the sum of 
$\beta$-decay lifetimes for the progenitor nuclei in this mass region.
Before reaching the progenitor nuclei with the $N=126$ closed neutron shell
that will produce the Pt peak after exhaustion of all the neutrons,
the nuclear flow must pass through the progenitor nuclei at $A\sim 130$
with the $N=82$ closed neutron shell. If the sum of $\beta$-decay
lifetimes for the $N=82$ progenitor nuclei is at least comparable to that
for the progenitor nuclei with $82< N\leq 126$, then the building-up of
the Pt peak is always controlled by the gradual diminishing of the
$N=82$ nuclei at $A\sim 130$, and a quasi-steady $\beta$-decay equilibrium
will result to largely determine the yield pattern in the 
$130< A\lesssim 195$ mass region (see, e.g., Kratz et al. 1993).
This quasi-steady $\beta$-decay equilibrium may well explain 
the observational result that the $r$-process elemental
abundance pattern for Ba and higher atomic numbers in very metal-poor stars
is remarkably close to the corresponding solar $r$-pattern.
Accordingly, we define $A > 130$ as the high mass region and $A\leq 130$
as the low mass region.
The lower end of the $A\leq 130$ region is not quite clear as it depends on,
e.g., the mass number of the seed nuclei for the $r$-process, but we will
assume it to be around $^{100}$Mo.

The solar $r$-process {\it nuclear} abundances 
(K\"appeler et al. 1989, 1991) are shown 
in terms of $\log\epsilon_{\odot,r}(A)$ in Figure 4(a).
The vertical dashed line in this figure indicates the
boundary ($A=130$) between the low and high mass regions chosen in our
simplified two-component model. 
As discussed above, we will use $Y_{127}^{\rm L}/Y_{127}^{\rm H}$ or
$Y_{182}^{\rm L}/Y_{182}^{\rm H}$ to fix the yield ratios 
$Y_{\cal{S}}^{\rm L}/Y_{\cal{S}}^{\rm H}$ for all the stable nuclei in the
$A\leq 130$ or $A > 130$ mass region, respectively.
Consequently, we can obtain
the yields of all the stable nuclei in the H and L events 
from the solar $r$-process {\it nuclear} abundances
shown in Figure 4(a) using 
the $\delta\log\epsilon_{\rm H}({\cal{S}})$ and
$\delta\log\epsilon_{\rm L}({\cal{S}})$ values for $^{127}$I and $^{182}$W
shown in Figure 3. However,
for comparison with stellar observations,
we have to use the elemental abundances. The solar $r$-process
{\it elemental} abundances in the low and high mass regions are shown
in terms of $\log\epsilon_{\odot,r}(Z)$
in Figure 4(b). As Xe (at atomic number $Z=54$)
has $r$-process isotopes in both the low and high mass regions, the vertical
dashed line indicating the boundary between the two mass regions
is put at $Z=54$ in Figure 4(b).
The elements Ba and Eu in the high ($A > 130$) mass region 
are commonly observed for very metal-poor stars
(e.g., McWilliam et al. 1995) and observations of Pd, Ag, and Cd 
in the low ($A\leq 130$) mass region 
are being actively pursued for such stars (Sneden et al. 1999).
The solar $r$-process abundances $N_{\odot,r}$ and the corresponding
$\log\epsilon_{\odot,r}$ values of these elements
are given in Table 2.
For comparison with observational data, we will focus on
Ag ($A\sim 107$) and Eu as the representative elements of 
the low and high mass regions, respectively.

In Figure 5, we show
$\log\epsilon_{\rm H}({\rm Ag})$, $\log\epsilon_{\rm L}({\rm Ag})$,
$\log\epsilon_{\rm H}({\rm Eu})$, and $\log\epsilon_{\rm L}({\rm Eu})$
as functions of $\Delta_{\rm L}$. As discussed above, 
these elemental yields are obtained from
the solar $r$-process {\it elemental} abundances shown in Figure 4(b)
with the use of the $\delta\log\epsilon_{\rm H}({\cal{S}})$ and
$\delta\log\epsilon_{\rm L}({\cal{S}})$ values for $^{127}$I and $^{182}$W
shown in Figure 3.
In scenario A [with the assumption of $\Delta_{\rm H}/\Delta_{\rm L}=0.1$,
cf. Fig. 5(a)], typical yields are
$\log\epsilon_{\rm L}({\rm Ag}) \approx -0.65$ and
$\log\epsilon_{\rm H}({\rm Eu}) \approx -2.38$ 
(with variations $\sim\pm 0.1$ and $\sim\pm 0.2$, respectively)
over the narrow physical range for $\Delta_{\rm L}$.
In addition, we have 
$\log\epsilon_{\rm H}({\rm Ag})\approx -3.3$ to $-2.7$ and
$\log\epsilon_{\rm L}({\rm Eu})\approx -2.4$ to $-2.0$ in 
regions of $\Delta_{\rm L}$ 
away from the limiting cases of 
$\Delta_{\rm L}^{\rm min}$ or
$\Delta_{\rm H}^{\rm max}$, respectively.
In scenario B [with the assumption of
$\Delta_{\rm H}/\Delta_{\rm L}=0.1$, cf. Fig. 5(b)], 
typical yields are
$\log\epsilon_{\rm L}({\rm Ag})\approx -0.64$,
$\log\epsilon_{\rm H}({\rm Eu})\approx-2.89$, and
$\log\epsilon_{\rm L}({\rm Eu})\approx -1.45$, with variations 
$\sim\pm(0.1$ to 0.2) over the range of
(1.06 to $2)\times 10^8$~yr for $\Delta_{\rm L}$.
Away from the limiting case of $\Delta_{\rm L}=\Delta_{\rm L}^{\rm min}$,
$\log\epsilon_{\rm H}({\rm Ag})$ also has a rather constant value 
$\sim -3.3$. The results for the specific case of 
$\Delta_{\rm L}=1.5\times 10^8$~yr and $\Delta_{\rm H}/\Delta_{\rm L}=0.1$
are typical of scenarios A and B, and are given in Table 3.

\subsection{Yield patterns and total mass yields
in the simplified two-component model}

Using the case of $\Delta_{\rm L}=1.5\times 10^8$~yr and
$\Delta_{\rm H}/\Delta_{\rm L}=0.1$ for both scenarios A and B as examples,
we show the yield patterns in the above simplified two-component model
in Figures 6 and 7. For comparison with stellar observations,
the elemental yields shown in these two figures are given
in terms of the $\log\epsilon$ values
for a star formed from an ISM
contaminated by a single H or L event. As in Figure 4,
the vertical dashed line at $Z=54$ (Xe) in Figures 6 and 7 indicates the
boundary ($A=130$) between the low and high mass regions in the 
simplified model. 
Note that the yield patterns in the H events are quite similar in
the chosen examples for scenarios A and B [cf. Figs. 6(a) and 7(a)]. 
The downward shifts in $\log\epsilon$ from scenario A to B 
are 0.3 and 0.5 dex
in the low and high mass regions, respectively.
However, while the yields in the low mass region in the L events
are essentially the same in these examples,
the yields in the high mass region
increase by 0.7 dex in $\log\epsilon$ from scenario A to B
[cf. Figs. 6(b) and 7(b)]. 

The dotted curve labelled ``TS''
in Figures 6 and 7 shows the solar $r$-pattern translated
to match the yield of Eu (hence those of $Z > 54$ by the assumption of
the simplified model) in the chosen examples. With respect to
the translated solar $r$-pattern in these examples for scenarios A and B,
the H event shows a depletion by 1.2 and 1.0 dex, respectively, while
the L event shows an enhancement by 0.8 and 0.1 dex, respectively,
of the yields below Xe.
Note that in the example for scenario B,
the gross yield pattern in the L event
almost coincides with the translated solar $r$-pattern.

To characterize the depletion of the yields in the low mass
region (below Xe) relative to those in the high mass region in the H event
with respect to the solar $r$-pattern, 
we define
\begin{equation}
\mbox{[Ag/Eu]}_r^{\rm H}\equiv
\log\left({Y_{\rm Ag}^{\rm H}\over Y_{\rm Eu}^{\rm H}}\right)-
\log\left[{N_{\odot,r}({\rm Ag})\over N_{\odot,r}({\rm Eu})}\right]
=\delta\log\epsilon_{\rm H}({\rm Ag})
-\delta\log\epsilon_{\rm H}({\rm Eu}).
\label{dep}
\end{equation}
In the simplified model, we have
$\delta\log\epsilon_{\rm H}({\rm Ag})=
\delta\log\epsilon_{\rm H}(^{127}{\rm I})$ and
$\delta\log\epsilon_{\rm H}({\rm Eu})=
\delta\log\epsilon_{\rm H}(^{182}{\rm W})$.
The values of [Ag/Eu]$_r^{\rm H}$ and the similarly defined 
[Ag/Eu]$_r^{\rm L}$ are given in Table 3 for the case of 
$\Delta_{\rm L}=1.5\times 10^8$~yr and
$\Delta_{\rm H}/\Delta_{\rm L}=0.1$ for both scenarios A and B.

We are also interested in comparing the total mass yields in the H and L
events. The ratio of the total
mass yield in an L event to that in an H event is
\begin{equation}
m\approx{\sum_{A=100}^{209}Y_A^{\rm L}\over\sum_{A=100}^{209}Y_A^{\rm H}},
\end{equation}
where we have neglected the small yields of the actinides ($A > 209$)
in both the H and L events.
For the specific examples shown in
Figures 6 and 7, we have $m\approx 12$ for scenario A 
[cf. Figs. 6(a) and (b)]
and $m\approx 58$
for scenario B [cf. Figs. 7(a) and (b)]. 
These values are typical of the two scenarios.
In \S5, we will briefly describe a possible mechanism that
can cause the total $r$-process yields from different supernova sources
to vary by factors $\gtrsim 10$ (cf. QVW98). However,
we note here that larger
differences in the total yields (e.g., $m\approx 58$)
may be more difficult to produce.

\section{Comparison with observed abundances in very metal-poor stars}

We now turn to the comparison of the yields in a single H or L event
in the above simplified two-component $r$-process model with the
observed abundances in very metal-poor stars using $\log\epsilon({\rm Eu})$
as a guide. In the examples where $\Delta_{\rm H}/\Delta_{\rm L}=0.1$ is
assumed, we have $\log\epsilon_{\rm H}({\rm Eu})=-2.62$ to $-2.22$ over
the entire physical range for $\Delta_{\rm L}$ and
$\log\epsilon_{\rm L}({\rm Eu})=-2.4$ to $-2.0$ over most of this range
in scenario A,
while in scenario B, we have $\log\epsilon_{\rm H}({\rm Eu})=-2.98$ to 
$-2.83$ and $\log\epsilon_{\rm L}({\rm Eu})=-1.62$ to $-1.30$ over
the range of (1.06 to $2)\times 10^8$~yr for $\Delta_{\rm L}$ (cf. Fig. 5).
The observed values of $\log\epsilon({\rm Eu})$ for four very metal-poor
stars (Sneden et al. 1996, 1998, 1999)
are given in Table 4. By comparing the $\log\epsilon({\rm Eu})$
values from our model with the observed ones, we can infer whether a
single H or L event in scenario A or B can explain the data reasonably well.
As indicated in Table 4, our inferences from such a comparison are:
(1) for HD 122563, a single H or L event in scenario A is
compatible with the observed $\log\epsilon({\rm Eu})$ value, 
while in scenario B, a single H event is 
only marginally compatible and a single L event 
is incompatible with the observation;
(2) for HD 115444, only a single L event
in scenario B is compatible with the observation; and (3) for CS 22892--052
with the lowest metallicity and HD 126238 with the highest metallicity of
the four stars shown in Table 4, we find that 
no single event is compatible with the data.

The comparison of the yields in a single event in our model with the 
observed abundances in HD 122563 is shown in Figure 8(a).
The dotted curve labelled ``TS'' in this figure is the solar $r$-pattern
translated to match the observed Eu abundance. In contrast, the solid
curve labelled ``H (A)'' or the dashed curve labelled ``L (A)'' shows
the yields in an H or L event in scenario A 
[cf. Figs. 6(a) or (b)] obtained directly from
our model (i.e., no fitting is attempted).
As we can see from Figure 8(a), the agreement between the solid curve
and the data is very good while the dashed curve describes the data
less well. However, a decisive test for the agreement between our
model and the data is the Ag abundance. It is predicted that 
the Ag abundance to be measured in HD 122563 should be
either higher by 0.8 dex [L (A)] or lower by over 1 dex [H (A)]
than the value corresponding to the translated solar $r$-pattern.
In Figure 8(b), the yields in an L event in scenario B obtained from
our model [cf. Fig. 7(b)] are shown by the solid curve labelled ``L (B),''
and can be compared with the observed abundances
in HD 115444. The solar $r$-pattern translated to match 
the observed Eu abundance (not shown)
is almost identical to the solid curve. We note 
the good agreement between the solid curve and the data, and emphasize
that a decisive test of this agreement is again
the Ag abundance to be measured in HD 115444.

It can be seen that our model does not describe the observations
very well (cf. Table 4) if the abundances in very 
metal-poor stars indeed represent
the result of a single supernova precursor contaminating the ISM with 
a constant dilution factor. It is also clear from the observations
alone that in order to account for the data,
a single supernova precursor model for the abundances in very metal-poor
stars has to invoke large variations in supernova production
of Fe and Eu, or otherwise
requires a grossly heterogeneous distribution of Fe relative to the heavy
$r$-process nuclei in the mixing of the supernova debris with the ISM. 
As shown in Table 4,
HD 122563 and HD 115444 have essentially the same 
$\log\epsilon({\rm Fe})$ value but differ 
in $\log\epsilon({\rm Eu})$ by about 1 dex, while CS22892--052 and
HD 126238 have essentially the same $\log\epsilon({\rm Eu})$ value but
differ in $\log\epsilon({\rm Fe})$
by about 1.4 dex. In addition, although CS22892--052 has 
the lowest $\log\epsilon({\rm Fe})$ value of the four stars 
shown in Table 4, its
$\log\epsilon({\rm Eu})$ value is higher than that of HD 122563
by about 1.5 dex. Using additional and more extensive data from 
McWilliam et al. (1995), we show
the $\log\epsilon({\rm Eu})$ and $\log\epsilon({\rm Fe})$ values for 15 very
metal-poor stars in Figure 9. Of these stars, two have [Fe/H] $=-2.06$ and
$-1.67$, respectively, and the rest have [Fe/H] $=-3.1$ to $-2.41$.
In contrast to the well-behaved Ba/Eu abundance ratio (close to the
corresponding solar $r$-process value, cf. McWilliam et al. 1995;
McWilliam 1998) shown in 
the upper part of Figure 9,
there is a large dispersion in $\log\epsilon({\rm Eu})$ over the range of
$\log\epsilon({\rm Fe})$ for the stellar sample with a corresponding 
large dispersion in the Eu/Fe abundance ratio. 

Using similar data and the framework of a single supernova precursor model,
Tsujimoto \& Shigeyama (1998) have attributed the large dispersion in
the Eu/Fe abundance ratio at very low metallicities to the different
dependences of Eu and Fe yields on the main sequence mass of the supernova
(see also Ishimaru \& Wanajo 1999). In addition, a stochastic chemical
evolution model for the early Galaxy
has been proposed to explain a similar dispersion in the Ba/Fe
abundance ratio (McWilliam 1997, 1998; McWilliam \& Searle 1999).
In this alternative model, individual regions are chemically enriched 
by random sampling of all possible supernova
yields through the occurrence of local supernovae.
Studies of the problem of Ba enrichment in the Galaxy by means of 
numerical simulations have been carried out by Raiteri et al. (1999)
(see also Travaglio et al. 1999).
These numerical studies attribute the large dispersion in the
Ba/Fe abundance ratio
at very low metallicities to the inhomogeneous chemical composition 
of the ISM from which the stars were formed.
While there may be numerous explanations for the large dispersion in
$\log\epsilon({\rm Eu})$ over the range of $\log\epsilon({\rm Fe})$
for very metal-poor stars, one possibility is that even at metallicities
of [Fe/H] $\sim -3.0$, the ISM may have already been contaminated by
more than a single supernova source. 
We will consider this possibility
using our  two-component $r$-process model in the following
discussion.

We first discuss the abundances of $r$-process elements 
in a star formed from an ISM
contaminated by a mixture of $n_{\rm H}$ H events and $n_{\rm L}$ L events.
In this case, the Ag and Eu abundances in the star are given by
\begin{equation}
10^{\log\epsilon({\rm Ag})}=
n_{\rm H}\times 10^{\log\epsilon_{\rm H}({\rm Ag})}+
n_{\rm L}\times 10^{\log\epsilon_{\rm L}({\rm Ag})},
\end{equation}
and
\begin{equation}
10^{\log\epsilon({\rm Eu})}=
n_{\rm H}\times 10^{\log\epsilon_{\rm H}({\rm Eu})}+
n_{\rm L}\times 10^{\log\epsilon_{\rm L}({\rm Eu})},
\end{equation}
respectively.
Let us focus on the case of $\Delta_{\rm L}=1.5\times 10^8$~yr and
$\Delta_{\rm H}/\Delta_{\rm L}=0.1$ for scenario A (cf. Table 3).
In order to account for the observed Eu abundance of
$\log\epsilon({\rm Eu})=-1.53$ in HD 115444 by pure H events, we need 
$n_{\rm H}=7$, which predicts an Ag abundance of
$\log\epsilon({\rm Ag})=-2.08$ for this star.
For the assumed values of $\Delta_{\rm H}$ and
$\Delta_{\rm L}$, the mixture of $n_{\rm H}=7$ and $n_{\rm L}=0$
would occur in a standard diluting mass
of the ISM with a Poisson probability
$P(n_{\rm H}=7,\ n_{\rm L}=0)=7.4\%$
over the first $1.05\times 10^8$~yr after the birth of the Galaxy.
We note that a second mixture of $n_{\rm H}=7$ and $n_{\rm L}=1$ would occur
equally likely in the same amount of ISM over the same period of time
with a Poisson probability $P(n_{\rm H}=7,\ n_{\rm L}=1)=5.2\%$.
While the second mixture gives $\log\epsilon({\rm Eu})=-1.44$ in
accord with the observed value in HD 115444, it predicts
a much higher Ag abundance of
$\log\epsilon({\rm Ag})=-0.63$ for this star.
The abundance pattern for the mixture of seven pure H events
is shown in Figure 10(a)
together with the data on HD 115444. The abundance pattern for the 
mixture of seven H events and one L event (not shown)
is almost identical to
the solar $r$-pattern translated to match the observed Eu abundance
(dotted curve labelled ``TS'' in this figure).

For CS 22892--052 and HD 126238, we first note that the observed Eu 
abundances are high compared with the yields in any single event
(see Tables 3 and 4).
We can account for the Eu data on these two stars 
with a mixture of 27 pure H events again using
the case of $\Delta_{\rm L}=1.5\times 10^8$~yr and
$\Delta_{\rm H}/\Delta_{\rm L}=0.1$ for scenario A. 
However, the value of 
$\log\epsilon({\rm Ag})=-1.50$ given by this mixture 
is far below the preliminary
value of $\log\epsilon({\rm Ag})\approx -0.75\pm0.25$ observed in
CS 22892--052 (Cowan \& Sneden 1999) while the comparison for HD 126238
cannot be made yet. On the other hand, the mixture of
$n_{\rm H}=26$ and $n_{\rm L}=1$, which
gives $\log\epsilon({\rm Ag})=-0.59$ and 
$\log\epsilon({\rm Eu})=-0.94$, fits the observations
of CS 22892--052,
and would occur in a standard diluting mass of the ISM with a Poisson 
probability $P(n_{\rm H}=26,\ n_{\rm L}=1)=1.5\%$ over 
the first $3.9\times 10^8$~yr after the birth of the Galaxy.
The abundance patterns for both the mixture of
$n_{\rm H}=27$ and $n_{\rm L}=0$ and that of 
$n_{\rm H}=26$ and $n_{\rm L}=1$ are
shown in Figure 10(b) together
with the data on CS 22892--052. Note that the preliminary Ag 
data clearly indicate a non-solar $r$-pattern in this star.
In addition to revealing possible
deviations of the $r$-pattern 
in very metal-poor stars from that in the solar system,
future observations of the abundances of Ag and other elements in the
low mass region in
HD 115444 and HD 126238 will further test 
whether mixtures of multiple
supernova $r$-process events in the context of our two-component model
may be a viable explanation. 

We next discuss the Fe abundance.
An estimate of the Fe yield in a single H or L event 
can be obtained under the assumption that
$Y_{\rm Fe}^{\rm H}\approx Y_{\rm Fe}^{\rm L}$ and
$\Delta_{\rm H}/\Delta_{\rm L}=0.1$. Following the derivation of the
$r$-process yields (eqs. [\ref{eh}] and [\ref{el}]), we obtain
\begin{equation}
\log\epsilon_{\rm H}({\rm Fe})\approx\log\epsilon_{\rm L}({\rm Fe})
\approx\log\epsilon_\odot({\rm Fe})+\log\alpha
-\log\left({T_{\rm UP}\over\Delta_{\rm H}}\right),
\end{equation}
where $\alpha$ is the fraction of Fe contributed by Type II supernovae
(the sources for the $r$-process nuclei in our model) to the solar
abundance of Fe 
[$\log\epsilon_\odot({\rm Fe})=7.51$, Anders \& Grevesse 1989]. 
For $\alpha\approx 1/3$ (cf. Timmes, Woosley, \& Weaver 1995)
and $\Delta_{\rm H}=1.5\times 10^7$~yr ($\Delta_{\rm L}=1.5\times 10^8$~yr),
we have $\log\epsilon_{\rm H}({\rm Fe})\approx
\log\epsilon_{\rm L}({\rm Fe})\approx 4.2$.
Inspection of Table 4 shows that except for CS 22892--052, all the three
other stars would require a mixture of supernova events to account for
their $\log\epsilon({\rm Fe})$ values. This contradicts the requirements
from the $\log\epsilon({\rm Eu})$ values in HD 122563 (single event) and
CS 22892--052 (mixture). For the mixtures given
above to account for the
$\log\epsilon({\rm Eu})$ values in HD 115444 ($n_{\rm H}+n_{\rm L}=7$ or 8)
and HD 126238 ($n_{\rm H}+n_{\rm L}=27$), we have
$\log\epsilon({\rm Fe})\approx 5.1$ (HD 115444) and 5.6 (HD 126238).
Agreement with the data is obtained only for HD 126238.
The poor representation of the data by a model assuming a constant
supernova Fe yield is not surprising. Theoretical estimates
(Timmes et al. 1995) show that the Fe yields are
very sensitive to the main sequence mass of the supernova 
at all metallicities, and 
may range from no to high production at zero metallicity.
Therefore, the ``metallicity'' as 
determined by [Fe/H] is at best only a rough guide to time in 
the early history of the Galaxy. It is not unreasonable for a star
such as CS 22892--052 with a very low [Fe/H] value to have been formed
from an ISM contaminated by $\sim 10$ supernova $r$-process events. 
We emphasize that if the timescale for replenishment of typical
molecular clouds with fresh $r$-process debris is $\sim 10^7$ yr, 
then the time resolution required to identify the earliest formed stars 
enriched in $r$-process elements is extremely fine 
($\Delta_{\rm H}/T_{\rm UP}\sim 10^{-3}$).
Therefore, the use of a rough chronometer such as [Fe/H] 
cannot define the relevant time periods.

From the remarkable regularity in the $r$-pattern for Ba and higher atomic
numbers observed in very metal-poor stars (Sneden et al. 1996, 1998, 1999),
it is reasonable to assume that the Eu 
abundance is a direct measure of the abundances in the high ($A > 130$) mass
region. Correspondingly, the Ag abundance may be taken as a direct measure
of the abundances in the low ($A\leq 130$) mass region.
With these assumptions, our two-component $r$-process model gives 
specific predictions for
$\log\epsilon_{\rm H}({\rm Eu})$ and $\log\epsilon_{\rm L}({\rm Ag})$,
and reasonably specific predictions for
$\log\epsilon_{\rm H}({\rm Ag})$ and $\log\epsilon_{\rm L}({\rm Eu})$ 
for both scenarios A and B.
This suggests that the proper measure of the age of very metal-poor stars
can be defined by $\log\epsilon({\rm Eu})$
rather than [Fe/H]. We infer that stars formed from an ISM
contaminated by a single
H event would have $\log\epsilon_{\rm H}({\rm Eu})=-2.98$ to $-2.22$ while
those formed from an ISM 
contaminated by a single L event would have the less restricted range
of $\log\epsilon_{\rm L}({\rm Eu})\approx -2.4$ to $-1.3$. To make
a specific assignment of an H or L event to a star will depend on 
concurrent measurements of abundances in both the low and high 
mass regions. As the H events are much more frequent, 
stars formed from an ISM contaminated by
a pure H event would be easier to find, while those formed from
an ISM contaminated by the 
less frequent L events should contain the debris from the H 
events that most likely preceded them.

Based on the typical values of $\log\epsilon_{\rm H}({\rm Eu})$ and
$\log\epsilon_{\rm L}({\rm Eu})$ given in Table 3 and
inspection of Table 4 and Figure 9, we infer that many of the 
very metal-poor stars studied so far would have to be 
assigned to formation from an ISM 
contaminated by multiple $r$-process events. 
For a constant supernova Fe yield,
the values of $\log\epsilon({\rm Fe})$ for these stars
also cannot be explained
by a single supernova precursor. In any case, the observed
values of $\log\epsilon({\rm Fe})$ are not correlated with those
of $\log\epsilon({\rm Eu})$ at very low metallicities.
However, as noted previously, theoretical estimates of supernova Fe yields
are widely variable and using [Fe/H] as a measure of 
the age of very metal-poor stars is highly problematic. 
To account for the observations with our two-component 
$r$-process model, we are obliged to consider that 
Fe yields are not strongly coupled 
with the production of the $r$-process elements in Type II supernovae 
at very low metallicities. It is also conceivable 
that there is an unidentified additional source of Fe at very early times.

\section{Conclusions}

We have shown the general consequences of a phenomenological 
two-component $r$-process model based on the $^{129}$I and $^{182}$Hf
abundances in the early solar system. This model assumes a standard
mass of the ISM for dilution of the debris from an individual supernova.
Two scenarios have been investigated to provide 
bounds on the model. 
The frequencies of the H ($\bar f_{\rm H}=1/\Delta_{\rm H}$) and L
($\bar f_{\rm L}=1/\Delta_{\rm L}$) events proposed in the model
are constrained by the meteoritic data on $^{129}$I and $^{182}$Hf.
The yields in a single H or L event are determined from these 
meteoritic data and the solar
$r$-process abundances under the assumption that 
the yield template in the low ($A\leq 130$) or high ($A > 130$) mass
region is the same for both the H and L events and follows the corresponding
solar $r$-pattern in each mass region.
These yields are represented by
the $\log\epsilon$ values for
a star formed from an ISM
contaminated by a single H or L event (cf. Figs. 6 and 7).
In this approach, the Eu abundance in a single 
H event is well defined (cf. Fig. 5). With the addition of subsequent 
supernova $r$-process debris to the ISM,
the abundances from further mixtures of multiple H and L events
can be obtained in a straightforward manner. This leads
to rather explicit quantitative predictions for stellar abundances
in the early Galaxy and for
the contrast between the stellar abundance
pattern at early times and the solar $r$-pattern. 
These predictions may be directly tested by 
comparison with the observed abundances in the low and high mass regions
in very metal-poor stars.

It is well known that [Fe/H] is not a reliable estimator of the Galactic age. 
Considering the observed Eu abundances, 
we find that even at very low metallicities of [Fe/H] $\sim -3.0$,
the ISM may have already been contaminated by many $r$-process
events. Therefore, we propose that the abundance of Eu be the criterion for 
identifying the earliest stars formed in the Galaxy. We predict that 
those stars with $\log\epsilon({\rm Eu})=-2.98$ to $-2.22$ were formed
from an ISM
contaminated most likely by a single H event within the first $\sim 10^7$~yr
of the Galactic history and should have an Ag/Eu abundance ratio less than
the corresponding solar $r$-process value by a factor of at least 10.
The crucial test for these predictions will again be the measurement of 
abundances in very metal-poor stars. We recognize that 
measurements at the low Eu abundances indicated here may pose
very difficult observational problems.

In this paper, the fundamental problems of the evolution of supernovae
and the sites of the $r$-process have not been addressed. 
Our approach has been purely phenomenological. If we assume that 
the production of
all the $r$-process nuclei is associated with a proto-typical 
general supernova event, then a scenario may be suggested that might 
unify the H and L events (cf. QVW98). We consider that a supernova
in its earlier stages ejects matter from the proto-neutron star
for the $r$-process with a relatively high number of neutrons per seed
nucleus (i.e., a relatively high neutron-to-seed ratio).
The $r$-process then dominantly produces nuclei in the high 
mass region with 
relatively few residual nuclei in the low mass region. These 
events are usually (i.e., at a high frequency) 
terminated by collapse of the proto-neutron star into a black 
hole, as may be described by the scenario of Brown \& Bethe (1994). On 
occasion (i.e., at a low frequency), collapse into a black 
hole does not occur and the ejection of matter for the $r$-process 
continues but with a lower neutron-to-seed ratio. The $r$-process
then dominantly produces nuclei in the low mass region with 
significant yields also
for nuclei in the high mass region. The total mass
yield of $r$-process nuclei in the rare events is much higher due to
the longer duration of mass ejection from the stable neutron star.	
The relative production of black 
holes to neutron stars in supernovae is $\sim 10:1$ to account for
the ratio of frequencies for the H and L events. However, 
the physics that might 
be responsible for the scenario sketched above
remains to be explored.

\acknowledgments

This work is dedicated to David Norman Schramm and is in the
spirit of excitement, hypotheses, and observation that typified
his approach. One of us remembers participating in the early wonders of
nuclear cosmochronology and the search for extinct nuclei
during his thesis. The other remembers the dense presentations and
mysteries of earlier nuclear cosmochronologic reports and the
interest and excitement of the new studies. The approach used
here seeks to follow that of previous scholars.
``The true method of experience first lights the candle
(by hypothesis), and then by means of the candle shows the
way, commencing as it does with experience duly ordered ...
and from it educing axioms (`first fruits,' provisional
conclusions), and from established axioms again new experiments ...
Experiment itself shall judge.'' --- Francis Bacon, {\it Novum Organum}
(1620)

We greatly appreciate the support by John Cowan and 
Christopher Sneden in freely providing us information on 
their work and in maintaining a continued level of interest 
in testing alternative models, however speculative. 
Discussions with Andrew McWilliam on abundances in 
very metal-poor stars were of considerable aid. We thank Petr Vogel
for comments on an earlier draft of the paper. This 
work was supported in part by the US Department of Energy under contract
W-7405-ENG-36 and grant DE-FG03-88ER-13851,
and by NASA under grant
NAG 5-4076, Caltech Division Contribution No. 8641(1032).
Y.-Z. Q. was supported by the J. Robert Oppenheimer Fellowship at
Los Alamos National Laboratory.

\clearpage

\appendix

\section{General discussion of scenarios A and B}

Without the assumption of $\Delta_{\rm H}/\Delta_{\rm L}=0.1$,
some general results for scenario A can still be obtained by
considering the restrictions 
$\Delta_{\rm H} < \Delta_{\rm H}^{\rm max}=1.85\times 10^7$~yr and
$\Delta_{\rm L} > \Delta_{\rm L}^{\rm min}=1.06\times 10^8$~yr
on equations (\ref{la2}) and (\ref{ha2}).
As $f(X_{129}^{\rm L}) > 0$ and
$f(X_{129}^{\rm H}) > f(\Delta_{\rm H}^{\rm max}/\bar\tau_{129})=0.646$,
equation (\ref{la2}) gives
\begin{equation}
1+\left({Y_{127}^{\rm L}\over Y_{127}^{\rm H}}\right)
\left({\Delta_{\rm H}\over\Delta_{\rm L}}\right) >
{f(X_{129}^{\rm H})\over C_{\rm I}} > {0.646\over  C_{\rm I}},
\end{equation}
which means that the fractions of $^{127}$I contributed by 
the H and L events to
the corresponding total solar $r$-process abundance
(cf. eqs. [\ref{fli}] and [\ref{fhi}])
are $F_r^{\rm H}(^{127}{\rm I}) < C_{\rm I}/0.646=0.068$ and
$F_r^{\rm L}(^{127}{\rm I}) > 0.932$, respectively.

For $\Delta_{\rm L} > \Delta_{\rm L}^{\rm min}=1.06\times 10^8$~yr,
we have $f(X_{182}^{\rm L}) < 2.35\times 10^{-3}\ll C_{\rm Hf}$.
Equation (\ref{ha2}) then gives
\begin{equation}
1+\left({Y_{182}^{\rm L}\over Y_{182}^{\rm H}}\right)
\left({\Delta_{\rm H}\over\Delta_{\rm L}}\right)\approx
{f(X_{182}^{\rm H})\over C_{\rm Hf}}\leq {1\over C_{\rm Hf}},
\end{equation}
which means that the fractions of $^{182}$W contributed by
the H and L events to
the corresponding total solar $r$-process abundance 
(cf. eqs. [\ref{fhw}] and [\ref{flw}]) are
$F_r^{\rm H}(^{182}{\rm W})\gtrsim C_{\rm Hf}=0.45$ and
$F_r^{\rm L}(^{182}{\rm W})\lesssim 0.55$, respectively.
Therefore, the L events account for
essentially all of the solar $r$-process $^{127}$I and the H events account
for more than 45\% of
the solar $r$-process $^{182}$W in scenario A.

Following the general discussion of scenario A presented above and that of
scenario B presented in \S2.2, we can obtain some general results
on the yields of Ag and Eu in the simplified two-component $r$-process model
without assuming that
$\Delta_{\rm H}/\Delta_{\rm L}=0.1$. These results are best represented by
$\log\epsilon_{\rm L}({\rm Ag})$, $\log\epsilon_{\rm H}({\rm Eu})$,
[Ag/Eu]$_r^{\rm L}$, and [Ag/Eu]$_r^{\rm H}$ for scenario A, and by
$\log\epsilon_{\rm L}({\rm Ag})$, $\log\epsilon_{\rm L}({\rm Eu})$,
$\log\epsilon_{\rm H}({\rm Eu})$, and [Ag/Eu]$_r^{\rm H}$ for scenario B.
From equations (\ref{eh})--(\ref{dep}), we have
\begin{equation}
\log\epsilon_{\rm L}({\rm Ag})=\log\epsilon_{\odot,r}({\rm Ag})
+\log F_r^{\rm L}({\rm Ag})-\log(T_{\rm UP}/\Delta_{\rm L}),
\label{lag}
\end{equation}
\begin{equation}
\log\epsilon_{\rm L}({\rm Eu})=\log\epsilon_{\odot,r}({\rm Eu})
+\log F_r^{\rm L}({\rm Eu})-\log(T_{\rm UP}/\Delta_{\rm L}),
\label{leu}
\end{equation}
\begin{equation}
\log\epsilon_{\rm H}({\rm Eu})=\log\epsilon_{\odot,r}({\rm Eu})
+\log F_r^{\rm H}({\rm Eu})-\log(T_{\rm UP}/\Delta_{\rm H}),
\label{heu}
\end{equation}
\begin{equation}
\mbox{[Ag/Eu]}_r^{\rm L}=\log [F_r^{\rm L}({\rm Ag})/F_r^{\rm L}({\rm Eu})],
\label{ldep}
\end{equation}
and
\begin{equation}
\mbox{[Ag/Eu]}_r^{\rm H}=\log [F_r^{\rm H}({\rm Ag})/F_r^{\rm H}({\rm Eu})].
\label{hdep}
\end{equation}
In the simplified model, we assume that
$F_r^{\rm L}({\rm Ag})=F_r^{\rm L}(^{127}{\rm I})$
[hence $F_r^{\rm H}({\rm Ag})=F_r^{\rm H}(^{127}{\rm I})$], and
$F_r^{\rm H}({\rm Eu})=F_r^{\rm H}(^{182}{\rm W})$
[hence $F_r^{\rm L}({\rm Eu})=F_r^{\rm L}(^{182}{\rm W})$].

In both scenarios A and B,
essentially all of the low mass nuclei are produced in the L events,
i.e., $F_r^{\rm L}({\rm Ag})=F_r^{\rm L}(^{127}{\rm I})\approx 1$, 
and we have
$\Delta_{\rm L} > \Delta_{\rm L}^{\rm min}=1.06\times 10^8$~yr.
Equation (\ref{lag}) then gives
\begin{equation}
\log\epsilon_{\rm L}({\rm Ag})\approx
-0.81+\log\left({\Delta_{\rm L}\over 10^8\ {\rm yr}}\right),
\end{equation}
where we have assumed $T_{\rm UP}=10^{10}$~yr 
(here and elsewhere in the paper).
This is the most general result of the model. For a reasonable range
of (1.06 to $2)\times 10^8$~yr for $\Delta_{\rm L}$, we obtain
$\log\epsilon_{\rm L}({\rm Ag})\approx -0.78$ to $-0.51$ for both
scenarios A and B.

In scenario A, we have 
$F_r^{\rm H}({\rm Eu})=F_r^{\rm H}(^{182}{\rm W})\approx 0.45$ to 1
and $\Delta_{\rm H} < \Delta_{\rm H}^{\rm max}=1.85\times 10^7$~yr.
Equation (\ref{heu}) then gives
\begin{equation}
\log\epsilon_{\rm H}({\rm Eu})\approx
\mbox{($-2.83$ to $-2.49$)}
+\log\left({\Delta_{\rm H}\over 10^7\ {\rm yr}}\right).
\end{equation}
For a reasonable range of (1 to $1.85)\times 10^7$~yr for $\Delta_{\rm H}$,
we obtain $\log\epsilon_{\rm H}({\rm Eu})\approx -2.83$ to $-2.22$.
In addition, we have 
$F_r^{\rm L}({\rm Ag})=F_r^{\rm L}(^{127}{\rm I}) > 0.932$,
$F_r^{\rm H}({\rm Ag})=F_r^{\rm H}(^{127}{\rm I}) < 0.068$,
and $F_r^{\rm L}({\rm Eu})=F_r^{\rm L}(^{182}{\rm W})\lesssim 0.55$.
Equations (\ref{ldep}) and (\ref{hdep}) then give
\begin{equation}
\mbox{[Ag/Eu]}_r^{\rm L} > 0.23,
\end{equation}
and
\begin{equation}
\mbox{[Ag/Eu]}_r^{\rm H} < -0.82.
\end{equation}

In scenario B, we have 
$F_r^{\rm L}({\rm Eu})=F_r^{\rm L}(^{182}{\rm W})\approx 0.55$ to 1 and
$\Delta_{\rm L} > \Delta_{\rm L}^{\rm min}=1.06\times 10^8$~yr.
Equation (\ref{leu}) then gives
\begin{equation}
\log\epsilon_{\rm L}({\rm Eu})\approx
\mbox{($-1.75$ to $-1.49$)}
+\log\left({\Delta_{\rm L}\over 10^8\ {\rm yr}}\right).
\end{equation}
For a reasonable range of (1.06 to $2)\times 10^8$~yr for $\Delta_{\rm L}$,
we obtain $\log\epsilon_{\rm L}({\rm Eu})\approx -1.72$ to $-1.19$.

We note that although there is no restriction on $\Delta_{\rm H}$
in scenario B, the yield of Eu in an H event is still well constrained by
equation (\ref{hb3}).
With $F_r^{\rm H}({\rm Eu})=F_r^{\rm H}(^{182}{\rm W})$,
this equation can be rewritten as
\begin{equation}
{1\over F_r^{\rm H}({\rm Eu})}
\left({T_{\rm UP}\over\Delta_{\rm H}}\right)\approx
\left({1\over C_{\rm Hf}}\right){T_{\rm UP}/\bar\tau_{182}\over
1-\exp(-\Delta_{\rm H}/\bar\tau_{182})},
\end{equation}
which gives
\begin{equation}
\log\epsilon_{\rm H}({\rm Eu})
\approx-2.72+\log[1-\exp(-\Delta_{\rm H}/\bar\tau_{182})].
\label{hb4}
\end{equation}
For $\Delta_{\rm H}\gtrsim 10^7$~yr, we obtain
$\log\epsilon_{\rm H}({\rm Eu})\approx -2.99$ to $-2.72$.
Similarly, equation (\ref{lb3}) can be
rewritten as
\begin{equation}
{1\over F_r^{\rm H}({\rm Ag})}
\left({T_{\rm UP}\over\Delta_{\rm H}}\right) >
\left({1\over C_{\rm I}}\right){T_{\rm UP}/\bar\tau_{129}\over
1-\exp(-\Delta_{\rm H}/\bar\tau_{129})},
\end{equation}
which gives
\begin{equation}
\log\epsilon_{\rm H}({\rm Ag})
< -2.81+\log[1-\exp(-\Delta_{\rm H}/\bar\tau_{129})].
\label{lb4}
\end{equation}
Combining equations (\ref{hb4}) and (\ref{lb4}), we obtain
$\log\epsilon_{\rm H}({\rm Ag})-\log\epsilon_{\rm H}({\rm Eu})< -0.09$,
which corresponds to
\begin{equation}
\mbox{[Ag/Eu]}_r^{\rm H} < -0.77.
\end{equation}

\clearpage

\figcaption{Typical characteristics of the H and L events in scenario A.
With the assumption of $\Delta_{\rm H}/\Delta_{\rm L}=0.1$ in this example,
the restrictions 
$\Delta_{\rm H} < \Delta_{\rm H}^{\rm max}=1.85\times 10^7$~yr and
$\Delta_{\rm L} > \Delta_{\rm L}^{\rm min}=1.06\times 10^8$~yr specify
a physical range of (1.06 to $1.85)\times 10^8$~yr for $\Delta_{\rm L}$.
(a) The relative yield of $^{127}$I to $^{182}$W in an H or L event 
with respect to the corresponding solar $r$-process abundance ratio,
measured by $U_{\rm H}$ or $U_{\rm L}$ 
(cf. eqs. [\protect\ref{uh}] and [\protect\ref{ul}]), respectively,
as a function of $\Delta_{\rm L}$.
(b) The fraction of $^{127}$I contributed by the L events
to the total solar $r$-process abundance of $^{127}$I, 
$F_r^{\rm L}(^{127}{\rm I})$ (cf. eq. [\protect\ref{fli}]),
and the fraction of
$^{182}$W contributed by the H events
to the total solar $r$-process abundance of $^{182}$W, 
$F_r^{\rm H}(^{182}{\rm W})$ (cf. eq. [\protect\ref{fhw}]),
as functions of $\Delta_{\rm L}$.
Note that $F_r^{\rm L}(^{127}{\rm I})=1$ and $F_r^{\rm H}(^{182}{\rm W})=1$
correspond to the limiting cases of 
$\Delta_{\rm L} = \Delta_{\rm L}^{\rm min}$ and
$\Delta_{\rm H} = \Delta_{\rm H}^{\rm max}$ (see \S2.1.1), respectively.}

\figcaption{Typical characteristics of the H and L events
in scenario B. As in Figure 1,
$\Delta_{\rm H}/\Delta_{\rm L}=0.1$ is assumed. In addition,
$\Delta_{\rm L}\leq 2\times 10^8$~yr is imposed in this example.
(a) See caption for Figure 1(a). Note that $U_{\rm L}=1.44$ to 1.26
over the range of (1.06 to $2)\times 10^8$~yr for $\Delta_{\rm L}$, i.e.,
the relative yields in the low and high mass regions in an L event
essentially follow the solar $r$-pattern.
(b) See caption for Figure 1(b). As the L events dominate the production
of both $^{127}$I and $^{182}$W, $F_r^{\rm L}(^{127}{\rm I})$ and
$F_r^{\rm L}(^{182}{\rm W})=1-F_r^{\rm H}(^{182}{\rm W})$ 
are shown as functions of $\Delta_{\rm L}$.
Note that $F_r^{\rm L}(^{127}{\rm I})=1$ corresponds to the
limiting case of $\Delta_{\rm L} = \Delta_{\rm L}^{\rm min}$ as in
Figure 1.}

\figcaption{(a) The yields of the stable (${\cal{S}}$) nuclei
$^{127}$I and $^{182}$W in an H or L event shown in terms of
the parameters
$\delta\log\epsilon_{\rm H}({\cal{S}})\equiv\log\epsilon_{\rm H}({\cal{S}})-
\log\epsilon_{\odot,r}({\cal{S}})$ or
$\delta\log\epsilon_{\rm L}({\cal{S}})\equiv\log\epsilon_{\rm L}({\cal{S}})-
\log\epsilon_{\odot,r}({\cal{S}})$ as functions of $\Delta_{\rm L}$
for scenario A. These parameters are calculated from the values of
$F_r^{\rm L}(^{127}{\rm I})$ and $F_r^{\rm H}(^{182}{\rm W})$
shown in Figure 1(b) with the use of equations (\protect\ref{eh}) 
and (\protect\ref{el}).
Note that $\log\epsilon_{\rm H}({\cal{S}})$ or
$\log\epsilon_{\rm L}({\cal{S}})$ gives the abundance of ${\cal{S}}$ nuclei
with respect to hydrogen in a star formed from an ISM contaminated by
a single H or L event, respectively. In standard spectroscopic notation,
$\log\epsilon({\cal{S}})=\log({\cal{S}}/{\cal{H}})+12$,
where ${\cal{S}}/{\cal{H}}$ is the abundance ratio of
the nuclide ${\cal{S}}$
to hydrogen in the star.
(b) Same as (a), but for scenario B. The values of
$F_r^{\rm L}(^{127}{\rm I})$ and $F_r^{\rm L}(^{182}{\rm W})$ shown
in Figure 2(b) are used in the calculation.}  

\figcaption{(a) The solar $r$-process nuclear abundances 
(K\"appeler et al. 1989, 1991) shown in
terms of $\log\epsilon_{\odot,r}(A)$. The vertical dashed line indicates
the boundary ($A=130$)
between the low and high mass regions chosen in our simplified
two-component $r$-process model.
(b) The solar $r$-process elemental abundances in the low and high mass
regions shown in terms of $\log\epsilon_{\odot,r}(Z)$. As in (a),
the vertical dashed line indicates the boundary between the two mass 
regions. This line is put at $Z=54$ (Xe) as Xe has $r$-process isotopes
in both mass regions.}

\figcaption{(a) The yields of Ag and Eu in an H or L event in scenario A
shown in terms of $\log\epsilon_{\rm H}({\rm Ag})$ and
$\log\epsilon_{\rm H}({\rm Eu})$ or $\log\epsilon_{\rm L}({\rm Ag})$ and
$\log\epsilon_{\rm L}({\rm Eu})$ as functions of $\Delta_{\rm L}$.
These $\log\epsilon$ values correspond to the abundances of Ag and Eu that 
should be observed in a star formed from an ISM contaminated by a single H 
or L event in scenario A.
They are obtained from the solar $r$-process elemental abundances
shown in Figure 4(b) with the use of the 
$\delta\log\epsilon_{\rm H}({\cal{S}})$ and
$\delta\log\epsilon_{\rm L}({\cal{S}})$ values for $^{127}$I and $^{182}$W
shown in Figure 3(a). Note that in our simplified two-component $r$-process 
model, we assume, for example,
$\delta\log\epsilon_{\rm L}({\rm Ag})=
\delta\log\epsilon_{\rm L}(^{127}{\rm I})$ and
$\delta\log\epsilon_{\rm H}({\rm Eu})=
\delta\log\epsilon_{\rm H}(^{182}{\rm W})$. 
(b) Same as (a), but for scenario B. The parameters shown in Figure 3(b)
are used to obtain these results.}

\figcaption{(a) The elemental yields in an H event in scenario A 
[solid curve labelled ``H (A)''] shown
in terms of $\log\epsilon(Z)$ for the case of
$\Delta_{\rm L}=1.5\times 10^8$~yr
and $\Delta_{\rm H}/\Delta_{\rm L}=0.1$. The dotted curve labelled ``TS''
is the solar $r$-pattern translated to match the
yield of Eu (hence those of $Z > 54$ by the assumption of our model).
The vertical line at Xe ($Z=54$) indicates the boundary between the low
and high mass regions chosen in our model. With respect to the translated
solar $r$-pattern, the yields below Xe are lower by 1.2 dex.
(b) The elemental yields in an L event in scenario A 
[solid curve labelled ``L (A)''] shown
in terms of $\log\epsilon(Z)$ for the same case as in (a).
Common symbols have the same
meanings as in (a). With respect to the translated solar $r$-pattern, 
the yields below Xe are higher by 0.8 dex.}

\figcaption{(a) The elemental yields in an H event in scenario B 
[solid curve labelled ``H (B)''] shown
in terms of $\log\epsilon(Z)$ for the case of
$\Delta_{\rm L}=1.5\times 10^8$~yr
and $\Delta_{\rm H}/\Delta_{\rm L}=0.1$. Common symbols have the same 
meanings as in
Figure 6. With respect to the translated solar $r$-pattern, the yields 
below Xe are lower by 1.0 dex.
(b) The elemental yields in an L event in scenario B 
[solid curve labelled ``L (B)''] shown
in terms of $\log\epsilon(Z)$ for the same case as in (a).
Common symbols have the same meanings as in Figure 6.
The yields in both the low and high mass regions
essentially coincide with the translated solar $r$-pattern.}

\figcaption{(a) Comparison of the yields in an H or L event
in scenario A 
with the observed abundances in HD 122563 given by
Sneden et al. (1998). The filled squares with error bars represent measured
abundances while the open triangles indicate upper limits.
The dotted curve labelled ``TS'' is the solar $r$-pattern
translated to match the observed Eu abundance (indicated by
``Eu'' below the filled square). In contrast, the solid
curve labelled ``H (A)'' or the dashed curve labelled ``L (A)'' shows
the yields in an H or L event in scenario A
[cf. Figs. 6(a) or (b)] obtained directly from
our model (i.e., no fitting is attempted). A decisive test of the
agreement between the model and the data is the abundances of elements
such as Ag in the low mass region, the observations of which are
under way (Sneden et al. 1999).
(b) Comparison of the yields in an L event in scenario B
with the observed abundances in HD 115444 given by
Sneden et al. (1998). Data symbols are the same as in (a).
The solid curve labelled ``L (B)'' shows the yields in an L event in
scenario B [cf. Fig. 7(b)] obtained directly from our model.
The solar $r$-pattern translated to match the observed Eu abundance
(not shown) is almost identical to the solid curve. As in (a),
the abundances in the low mass region 
being observed by Sneden et al. (1999)
will test decisively
the agreement between the model and the data.}

\figcaption{The observed Eu abundances and Ba/Eu abundance ratios
for 15 very metal-poor stars shown against the corresponding
Fe abundances (or metallicities [Fe/H]). The open squares represent
data from
McWilliam et al. (1995) and McWilliam (1998), while the filled triangles
represent data 
from Sneden et al. (1996, 1998, 1999). The quantity [Ba/Eu]$_r$ is
defined as [Ba/Eu]$_r\equiv\log({\rm Ba/Eu})-
\log[N_{\odot,r}({\rm Ba})/N_{\odot,r}({\rm Eu})]$, 
and measures the deviation
of the Ba/Eu abundance ratio from the corresponding
solar $r$-process value.
Note that except for one star, all the other stars with [Fe/H] $\leq -2.41$ 
shown in this figure have
[Ba/Eu]$_r\approx 0$ to 0.3. In contrast, the Eu abundances vary by about
1.5 dex for the same group of stars.}

\figcaption{(a) Comparison of the result from a mixture of
multiple $r$-process events
with the observed abundances in HD 115444 given by Sneden et al. (1998).
Common symbols have the same meanings as in Figure 8. 
The number of pure H events in
scenario A [cf. Fig. 6(a)] required to account for the observed Eu 
abundance is seven. The abundances from such a mixture are shown
by the solid curve labelled ``7 H (A),'' which is far below the
translated solar $r$-pattern in the low mass region. The result from
a mixture of seven H events and one L event [cf. Fig. 6(b)] in scenario A
(not shown) essentially coincides with the translated
solar $r$-pattern. Observations of abundances in the
low mass region will provide
a crucial test for the viability of using mixtures of
multiple $r$-process events to explain the data in HD 115444.
(b) Comparison of results from mixtures of
multiple $r$-process events
with the observed abundances in CS 22892--052 given by 
Sneden et al. (1996, 1999).
Common symbols have the same meanings as in Figure 8. 
The number of pure H events in
scenario A [cf. Fig. 6(a)] required to account for the observed Eu
abundance is 27. The abundances from such a mixture are shown
by the dashed curve labelled ``27 H (A),'' which is far below the
translated solar $r$-pattern in the low mass region. This mixture
cannot account for the preliminary value of the observed Ag abundance
(Cowan \& Sneden 1999) indicated by the open square. However,
the result from a mixture of 26 H events and one L event [cf. Fig. 6(b)]
in scenario A, shown as the solid curve labelled ``26 H (A) + 1 L (A),''
matches both the preliminary Ag data and 
the observed abundances
in the high mass region.}

\clearpage

\begin{deluxetable}{llc}
\footnotesize
\tablecaption{Input Data for the Two-Component $r$-Process Model}
\tablehead{
\colhead{$^{129}$I} & \colhead{$^{182}$Hf} & \colhead{Note}
}
\startdata
$\bar\tau_{129}=2.27\times 10^7$~yr & 
$\bar\tau_{182}=1.30\times 10^7$~yr & \tablenotemark{(a)}\nl
$(^{129}{\rm I}/{^{127}{\rm I}})_{\rm SSF}
=10^{-4}$ &
$(^{182}{\rm Hf}/{^{180}{\rm Hf}})_{\rm SSF}
=2.4\times 10^{-4}$ & \tablenotemark{(b)}\nl
$N_\odot(^{127}{\rm I})=0.90$ & 
$N_{\odot}(^{180}{\rm Hf})=0.0541$ & \tablenotemark{(c)}\nl
$N_{\odot,r}(^{127}{\rm I})\approx N_\odot(^{127}{\rm I})$ &
$N_{\odot,r}(^{182}{\rm W})=0.0222$ & \tablenotemark{(d)}\nl
$C_{\rm I}=0.0441$ &
$C_{\rm Hf}=0.450$ & \tablenotemark{(e)}\nl
\enddata
\tablenotetext{(a)}{Lifetimes of $^{129}$I and $^{182}$Hf.}
\tablenotetext{(b)}{Meteoritic data on $^{129}$I (Reynolds 1960; Jeffrey \&
Reynolds 1961; Brazzle et al. 1999) and $^{182}$Hf (Harper \& Jacobsen 1996;
Lee \& Halliday 1995, 1996, 1997, 1999).}
\tablenotetext{(c)}{Solar abundances of $^{127}$I and $^{180}$Hf on the scale
of $N_{\odot}({\rm Si})=10^6$ (Anders \& Grevesse 1989).}
\tablenotetext{(d)}{Solar $r$-process abundances of $^{127}$I 
(K\"appeler et al. 1989) and $^{182}$W (K\"appeler et al. 1991; 
Arlandini et al. 1999)
on the scale of $N_{\odot}({\rm Si})=10^6$.}
\tablenotetext{(e)}{Defined in
eqs. (\protect\ref{ci}) and (\protect\ref{chf}) and evaluated
for $Y_{129}/Y_{127}=1$ ($C_{\rm I}$) and $T_{\rm UP}=10^{10}$~yr
($C_{\rm I}$ and $C_{\rm Hf}$).}
\end{deluxetable}

\clearpage

\begin{deluxetable}{lllll}
\footnotesize
\tablecaption{Some Solar $r$-Process Elemental Abundance Data}
\tablehead{
\colhead{Element} & \colhead{$Z$\tablenotemark{(a)}} &
\colhead{$A$\tablenotemark{(b)}} & 
\colhead{$N_{\odot,r}$\tablenotemark{(c)}} &
\colhead{$\log\epsilon_{\odot,r}$\tablenotemark{(d)}}
}
\startdata
Pd & 46 & 105, 106, 108, 110 & 0.770 & 1.44\nl
Ag & 47 & 107, 109 & 0.435 & 1.19\nl
Cd & 48 & 111--114, 116 & 0.771 & 1.44\nl
Ba & 56 & 135, 137, 138 & 0.726\tablenotemark{(e)} & 1.42\nl
Eu & 63 & 151, 153 & 0.0907 & 0.51\nl
\enddata
\tablenotetext{(a)}{Atomic number.}
\tablenotetext{(b)}{Mass numbers of $r$-process isotopes.}
\tablenotetext{(c)}{Solar $r$-process abundance (K\"appeler et al. 1989)
on the scale of $N_{\odot}({\rm Si})=10^6$.}
\tablenotetext{(d)}{In standard spectroscopic notation,
$\log\epsilon_{\odot,r}=\log[N_{\odot,r}/N_\odot({\cal{H}})]+12$, where
$N_\odot({\cal{H}})=2.79\times 10^{10}$ is the solar abundance of hydrogen
(Anders \& Grevesse 1989) on the scale of $N_{\odot}({\rm Si})=10^6$.}
\tablenotetext{(e)}{Assumed to be $3\times N_{\odot,r}(^{135}{\rm Ba})$ 
in view of the uncertainties
in the $s$-process abundances (cf. K\"appeler et al. 1989;
Arlandini et al. 1999).}
\end{deluxetable}

\clearpage

\begin{deluxetable}{lllllll}
\footnotesize
\tablecaption{Typical Yields in the Simplified 
Two-Component $r$-Process Model}
\tablehead{
\colhead{Scenario\tablenotemark{(a)}} & 
\colhead{$\log\epsilon_{\rm H}({\rm Ag})$\tablenotemark{(b)}} &
\colhead{$\log\epsilon_{\rm H}({\rm Eu})$\tablenotemark{(b)}} &
\colhead{[Ag/Eu]$_r^{\rm H}$\tablenotemark{(c)}} &
\colhead{$\log\epsilon_{\rm L}({\rm Ag})$\tablenotemark{(b)}} &
\colhead{$\log\epsilon_{\rm L}({\rm Eu})$\tablenotemark{(b)}} &
\colhead{[Ag/Eu]$_r^{\rm L}$\tablenotemark{(c)}}
}
\startdata
A & $-2.93$ & $-2.38$ & $-1.22$ & $-0.65$ & $-2.13$ & 0.80 \nl
B & $-3.22$ & $-2.89$ & $-1.01$ & $-0.64$ & $-1.45$ & 0.12 \nl
\enddata
\tablenotetext{(a)}{For the case of $\Delta_{\rm L}=1.5\times 10^8$~yr and
$\Delta_{\rm H}/\Delta_{\rm L}=0.1$.}
\tablenotetext{(b)}{The yields of Ag and Eu in an H or L event in terms of 
the $\log\epsilon$ values for a star formed from an ISM
contaminated by an H or L event. In standard spectroscopic notation,
$\log\epsilon({\rm Ag})=\log({\rm Ag}/{\cal{H}})+12$, where
${\rm Ag}/{\cal{H}}$ is the abundance ratio of Ag 
to hydrogen in the star.}
\tablenotetext{(c)}{A measure for the relative yield of Ag to Eu in an
H or L event with respect to the corresponding solar $r$-process 
abundance ratio. See eq. (\protect\ref{dep}) for definition.}
\end{deluxetable}

\clearpage

\begin{deluxetable}{llllll}
\footnotesize
\tablecaption{Comparison with Observational Data on Very Metal-Poor Stars}
\tablehead{
\colhead{Star\tablenotemark{(a)}} & \colhead{[Fe/H]\tablenotemark{(b)}} &
\colhead{$\log\epsilon({\rm Fe})$\tablenotemark{(c)}} &
\colhead{$\log\epsilon({\rm Eu})$\tablenotemark{(d)}} &
\colhead{Single H Event?\tablenotemark{(e)}} &
\colhead{Single L Event?\tablenotemark{(e)}}
}
\startdata
HD 122563 & $-2.71$ & 4.8 & $-2.48$ & A (yes), B (?) & A (yes)\nl
HD 115444 & $-2.77$ & 4.7 & $-1.53$ & no & B (yes)\nl
CS 22892--052 & $-3.1$ & 4.4 & $-0.94$ & no & no\nl
HD 126238 & $-1.67$ & 5.8 & $-0.92$ & no & no\nl
\enddata
\tablenotetext{(a)}{Data on CS 22892--052 from Sneden et al. (1996, 1999), 
and the rest from Sneden et al. (1998).}
\tablenotetext{(b)}{Metallicity.}
\tablenotetext{(c)}{Calculated from 
$\log\epsilon({\rm Fe})=\log\epsilon_\odot({\rm Fe})+\mbox{[Fe/H]}$
with $\log\epsilon_\odot({\rm Fe})=7.51$ (Anders \& Grevesse 1989).
Note that for uniform Fe production by Type II supernovae, the yield
in an H or L event corresponds to
$\log\epsilon({\rm Fe})\approx 4.2$.}
\tablenotetext{(d)}{In standard spectroscopic notation,
$\log\epsilon({\rm Eu})=\log({\rm Eu}/{\cal{H}})+12$, where
${\rm Eu}/{\cal{H}}$ is the abundance ratio of Eu to 
hydrogen in the star.} 
\tablenotetext{(e)}{``A'' and ``B'' stand for scenarios A and B, 
respectively. A question mark (?) indicates a marginal case. 
Note that for both scenarios A and B, the expected range of Eu yield 
corresponds to
$\log\epsilon({\rm Eu})=-2.98$ to $-2.22$ for an H event
and $\log\epsilon({\rm Eu})\approx -2.4$ to $-1.3$ for an L event.
See Table 3 for typical Eu yields.} 
\end{deluxetable}

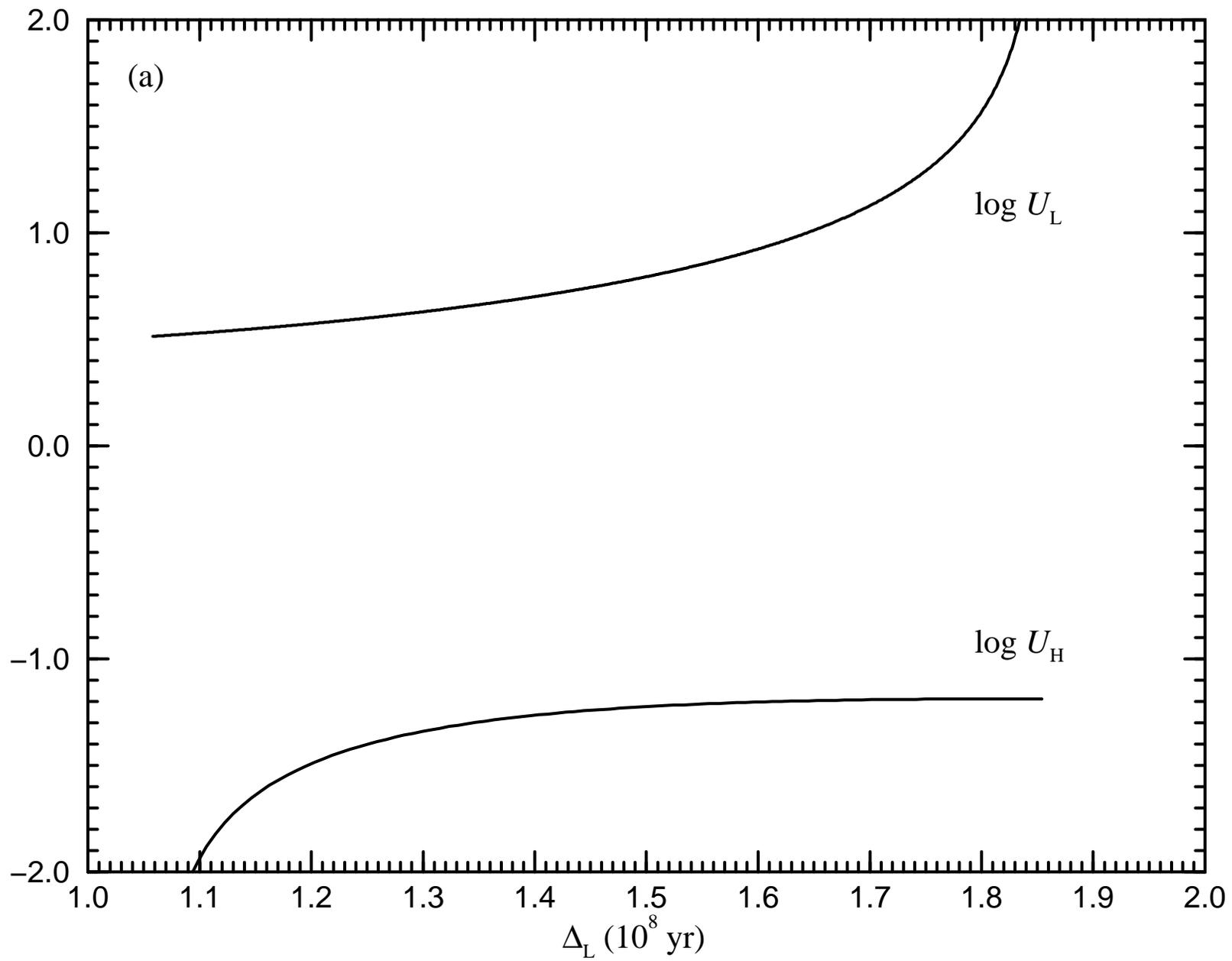
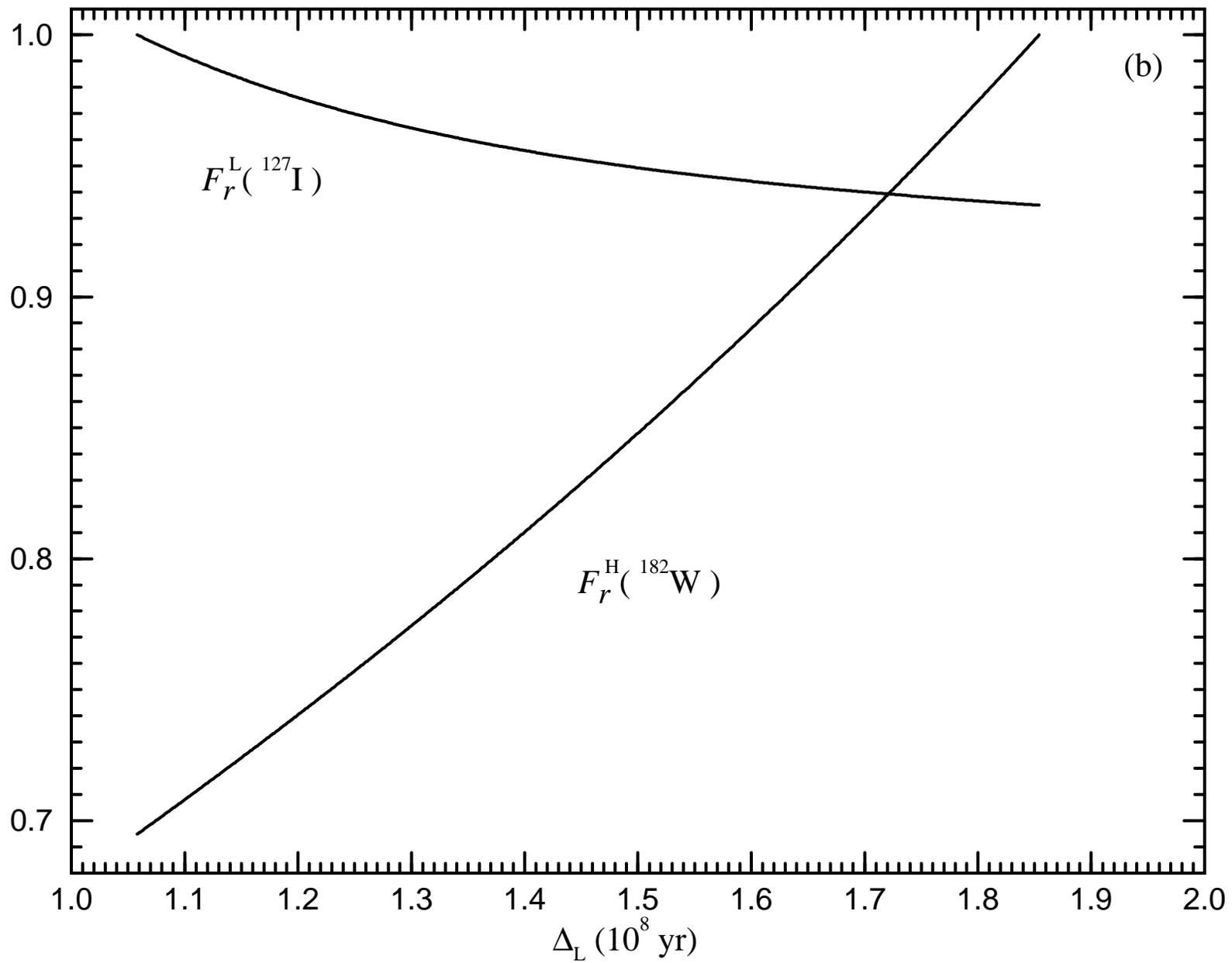
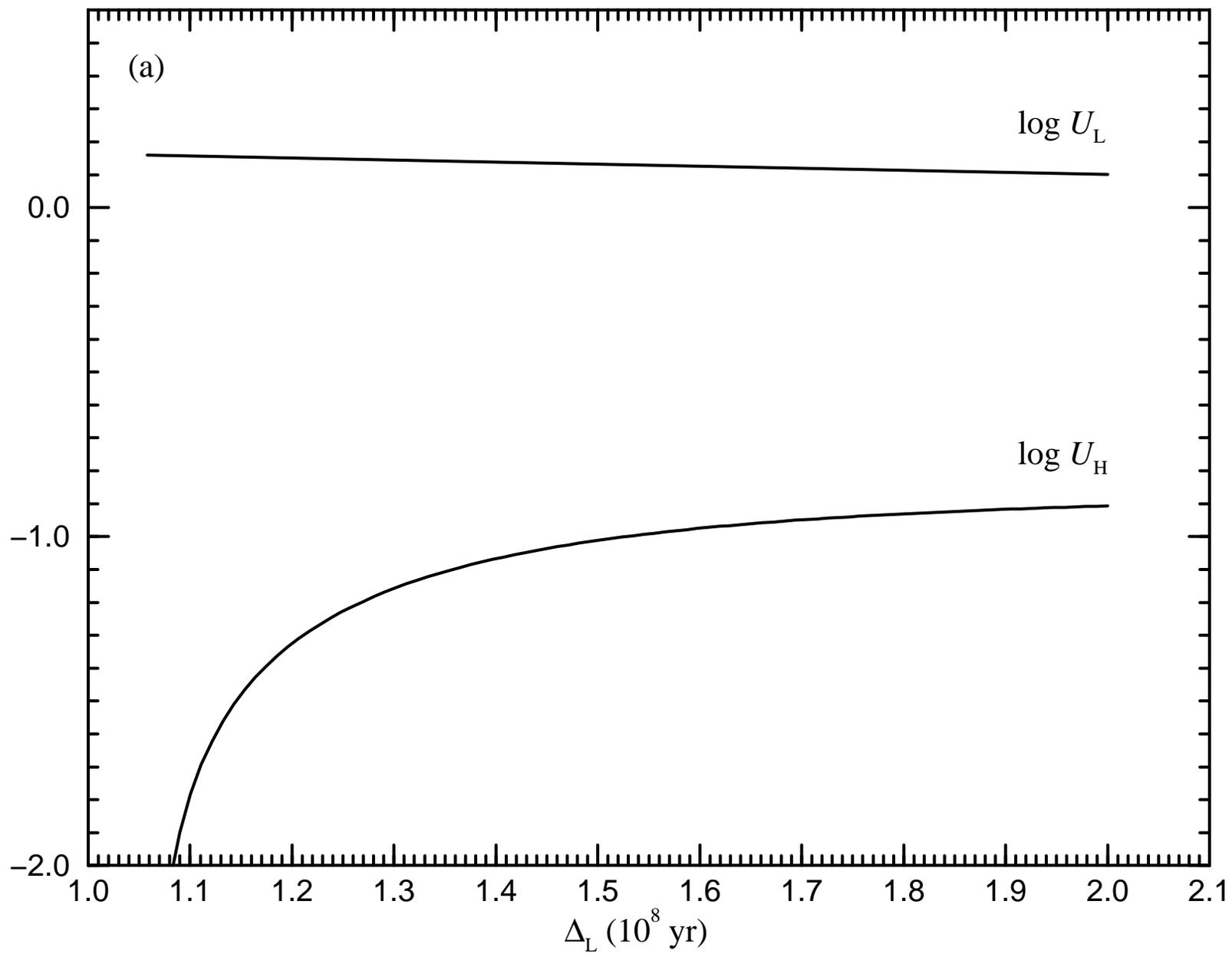
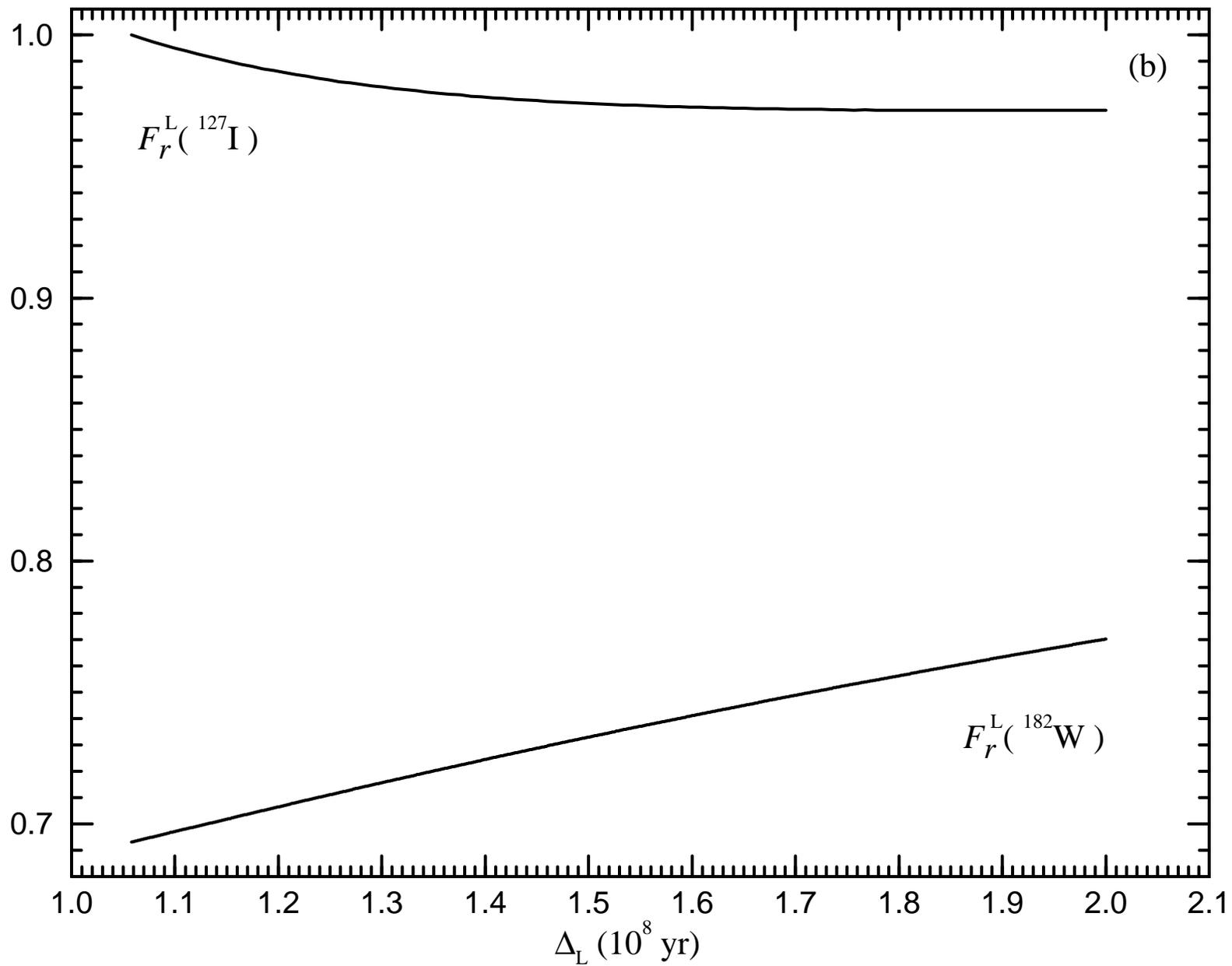
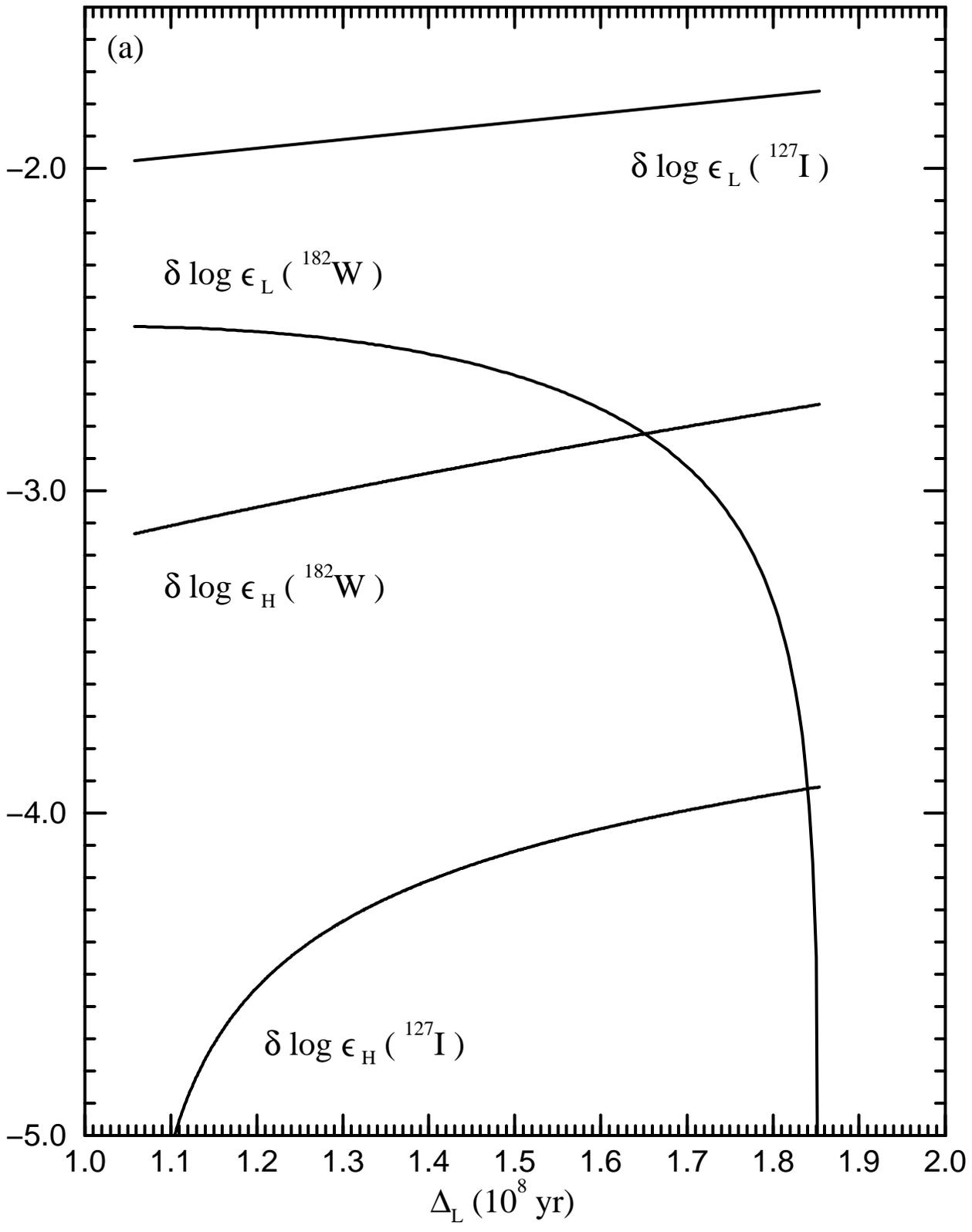
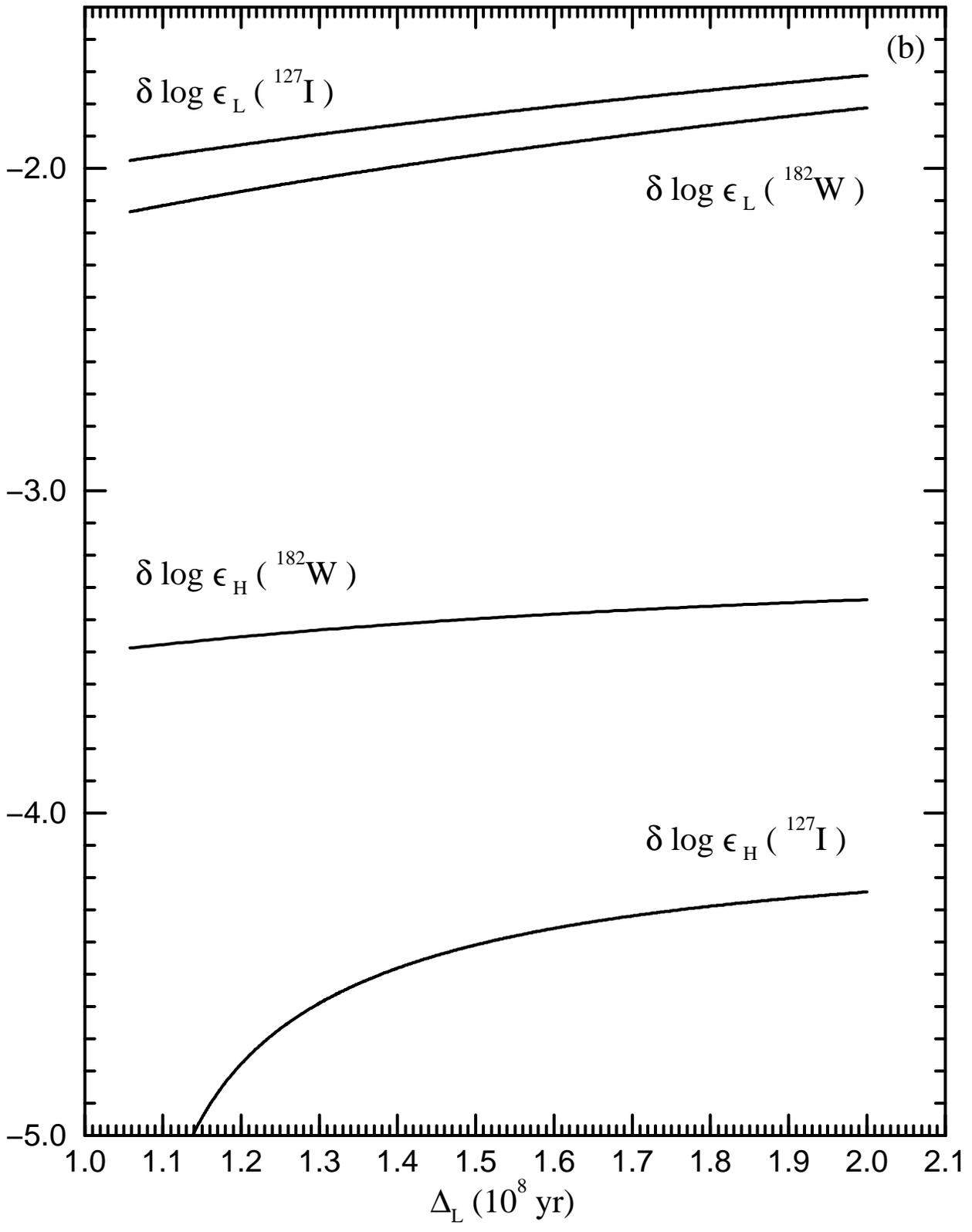
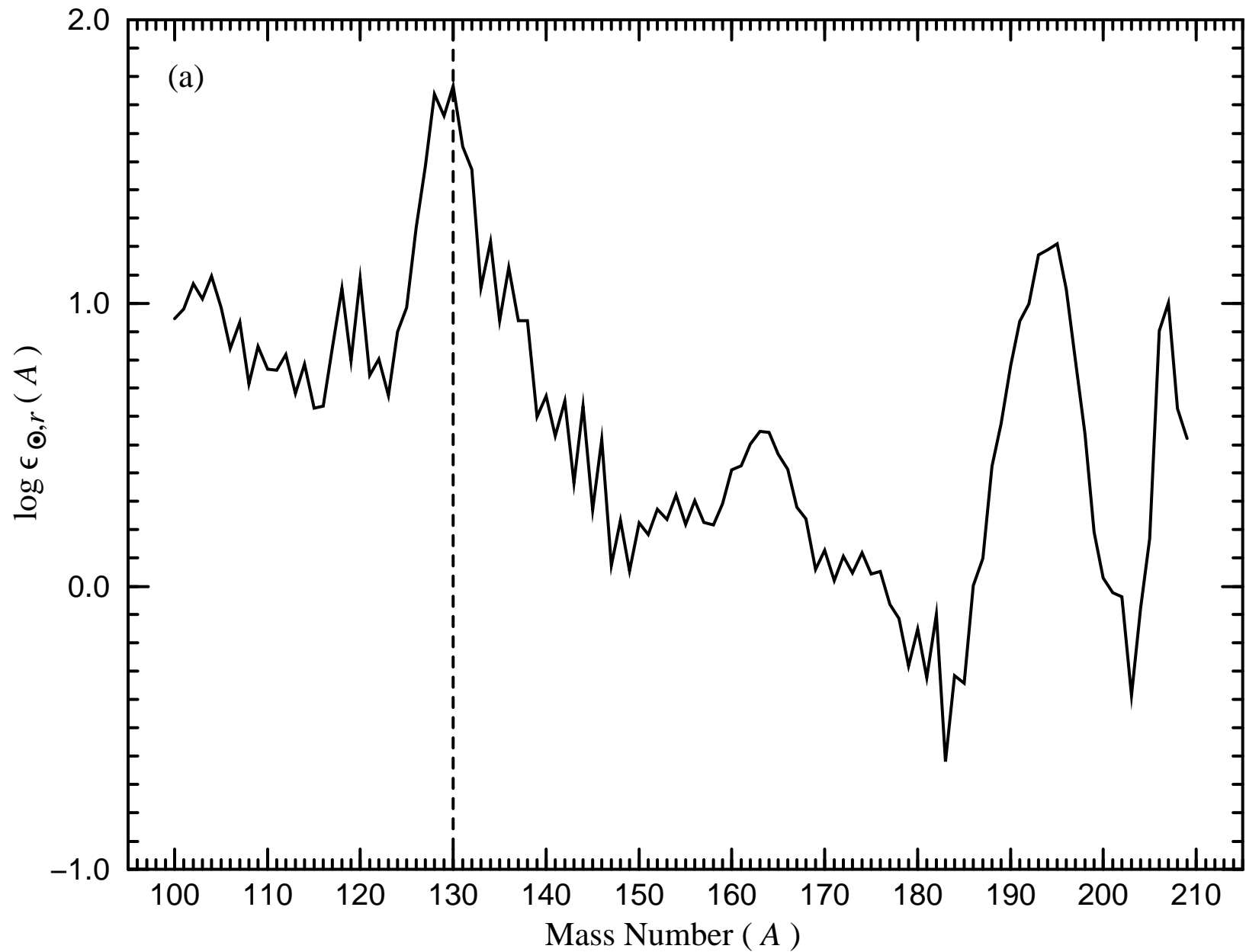
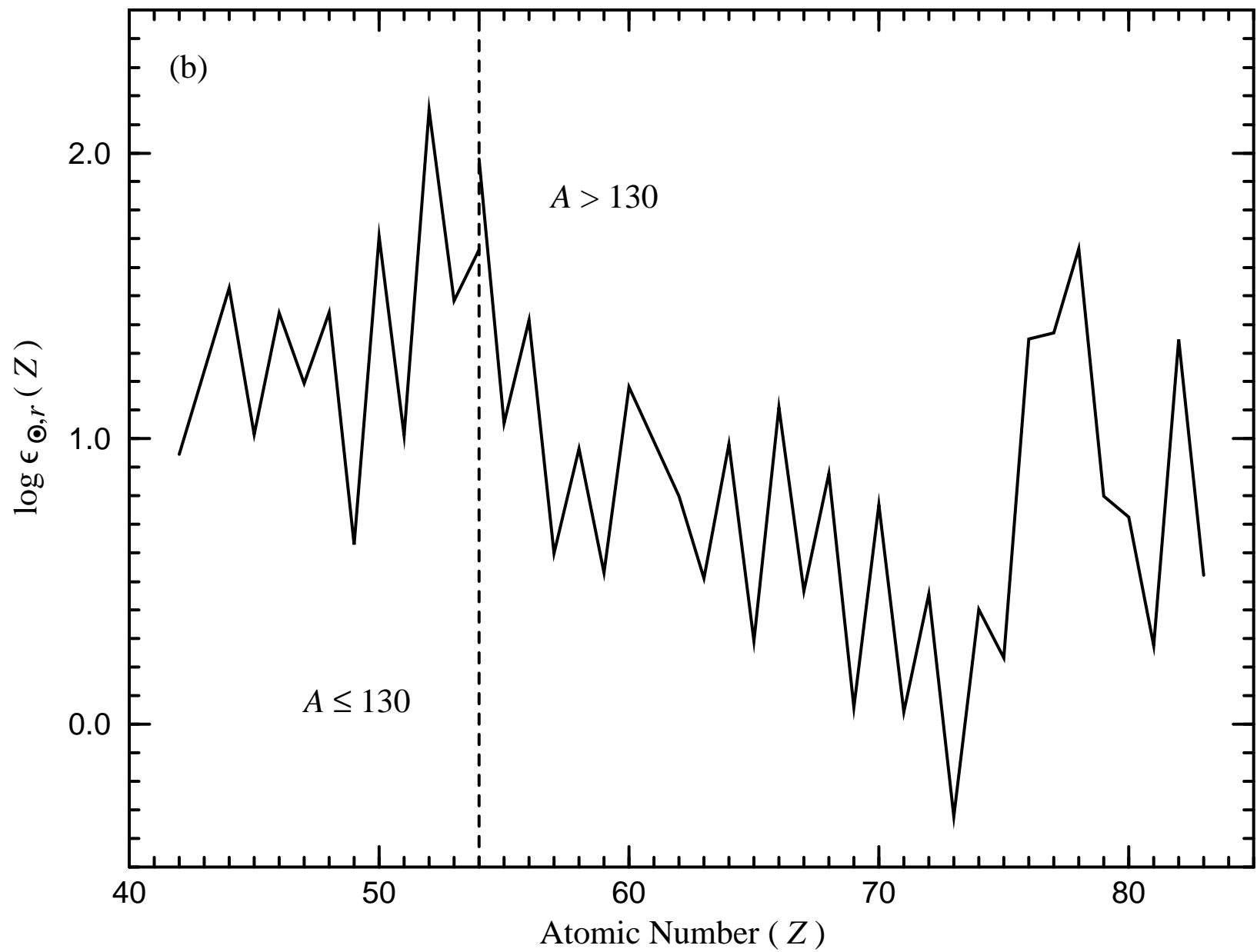
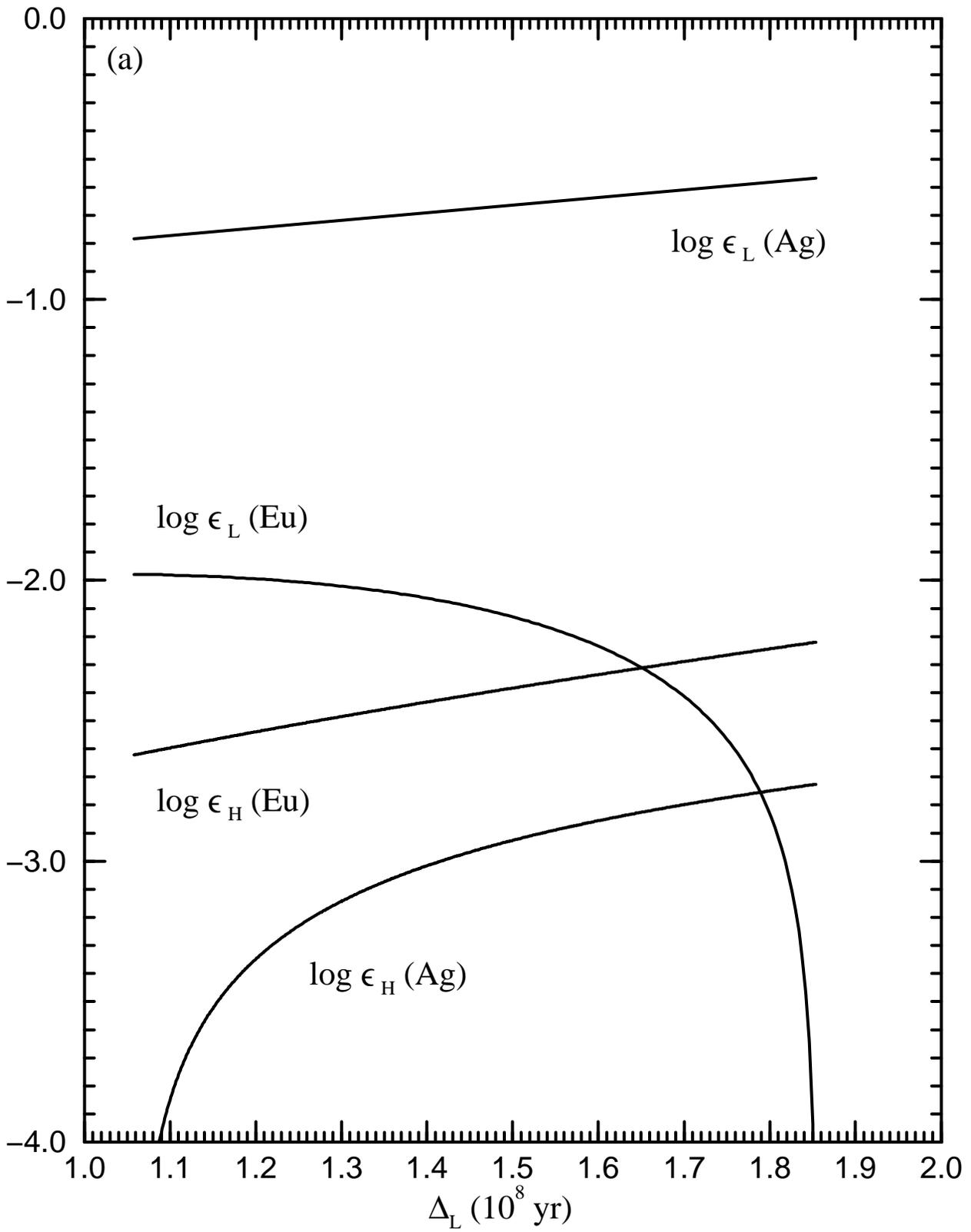
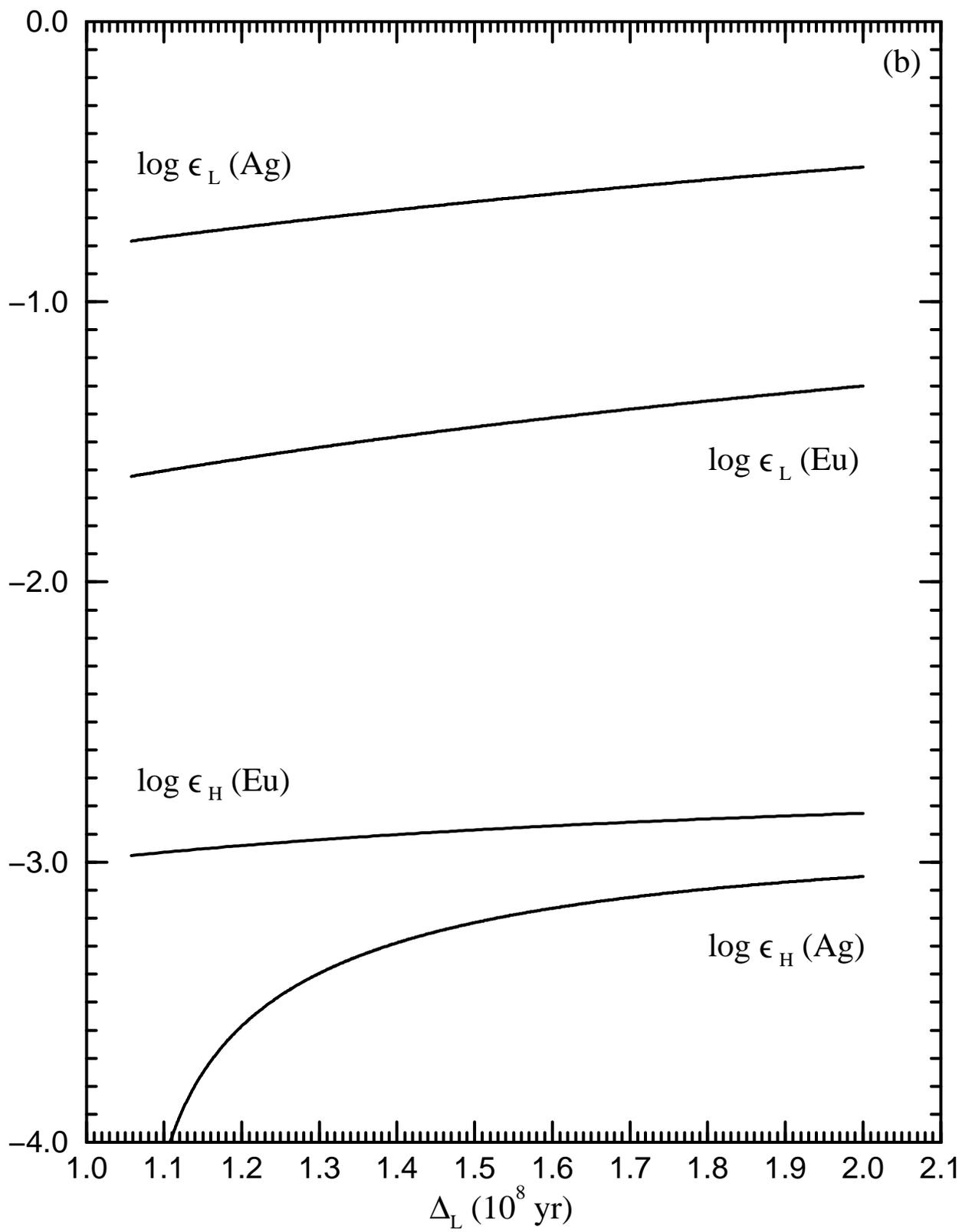
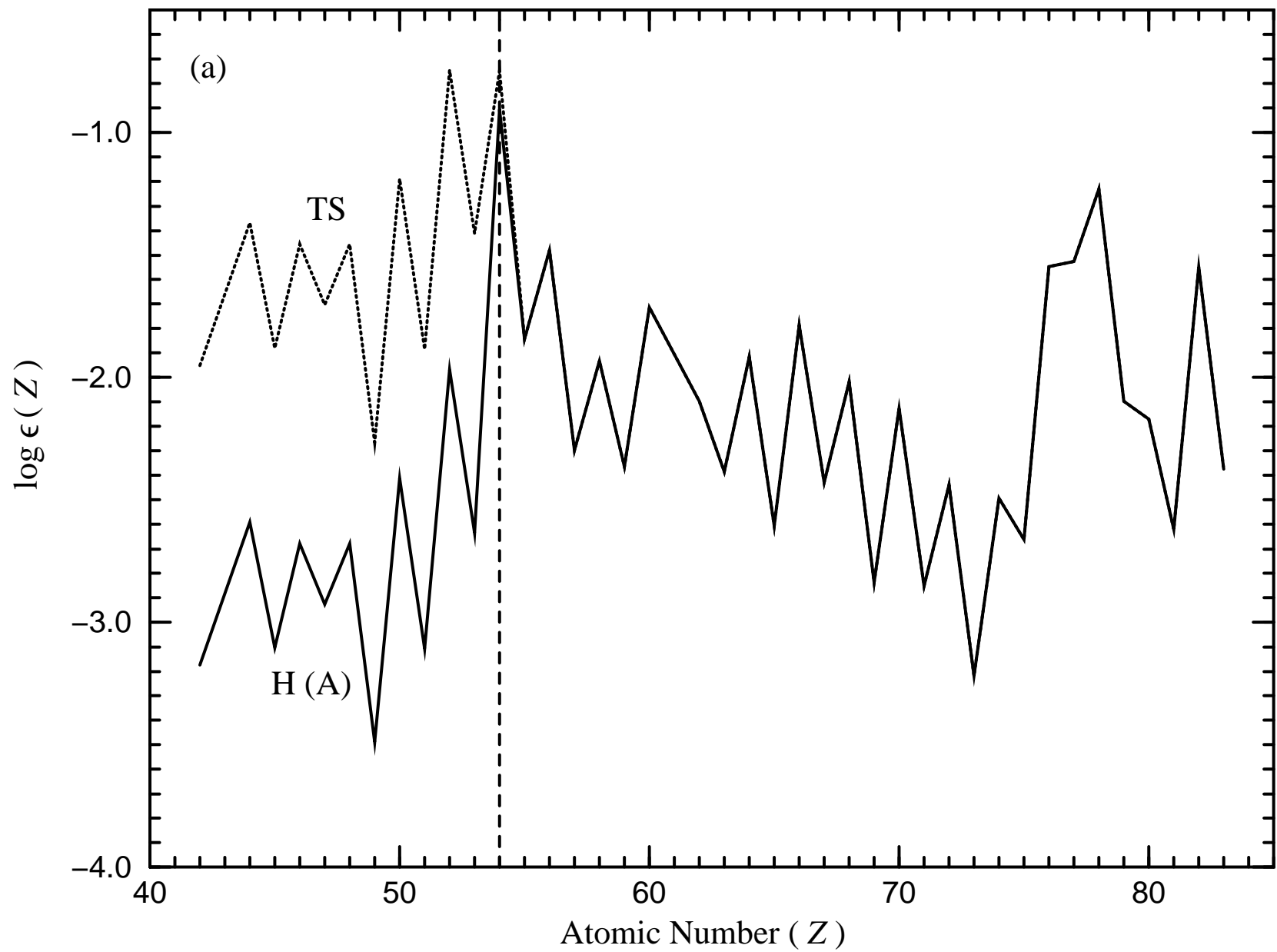
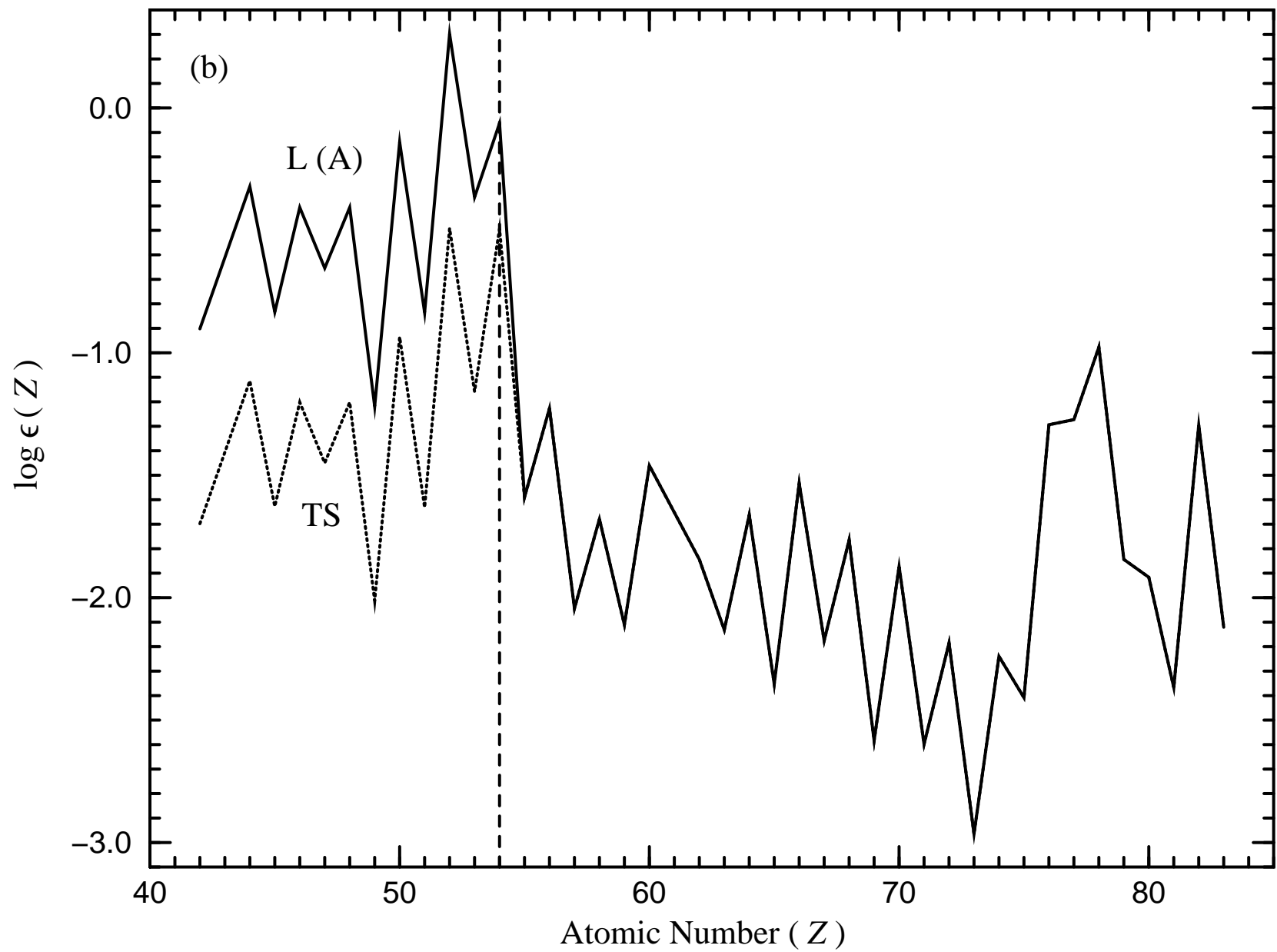
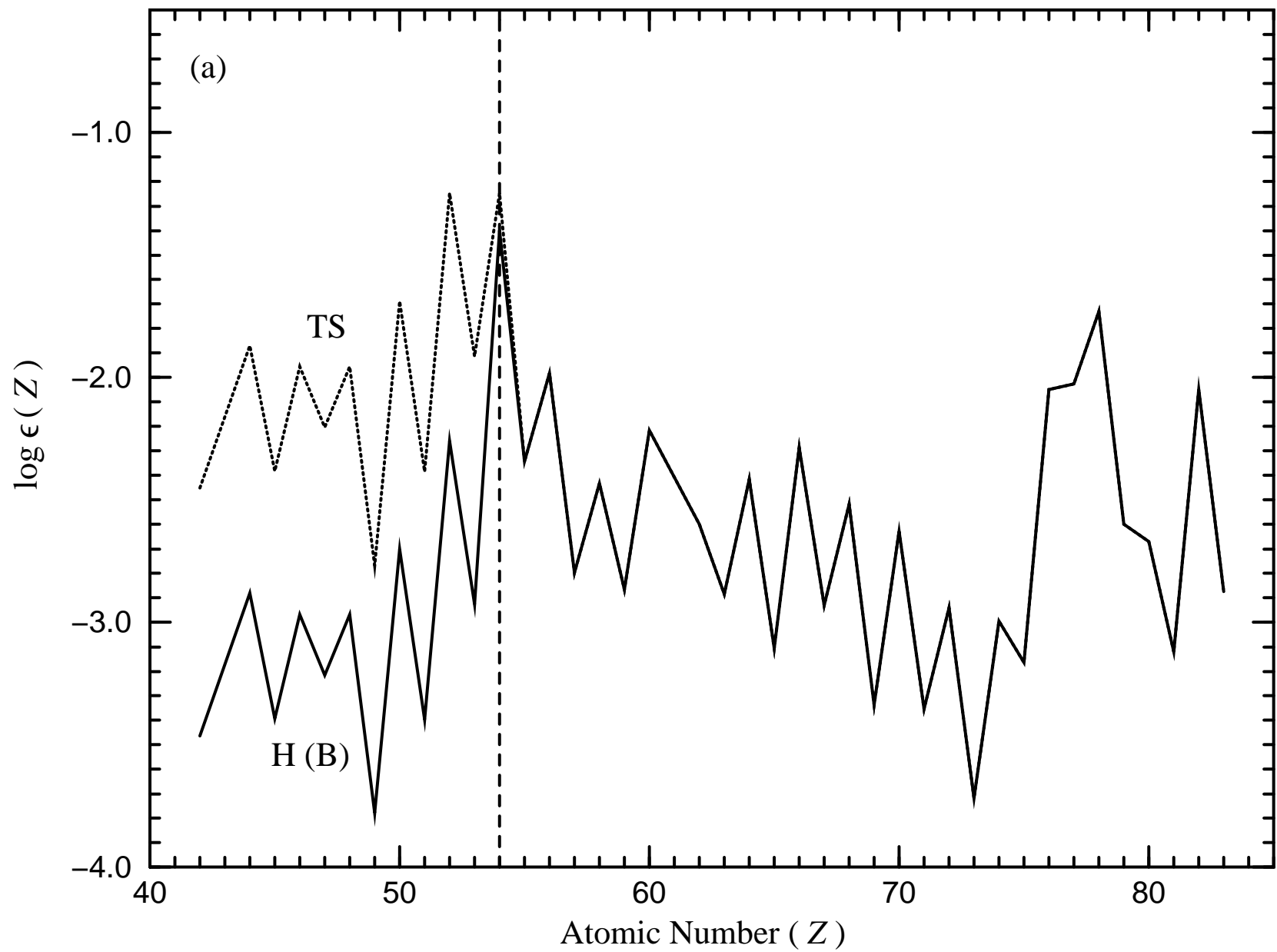
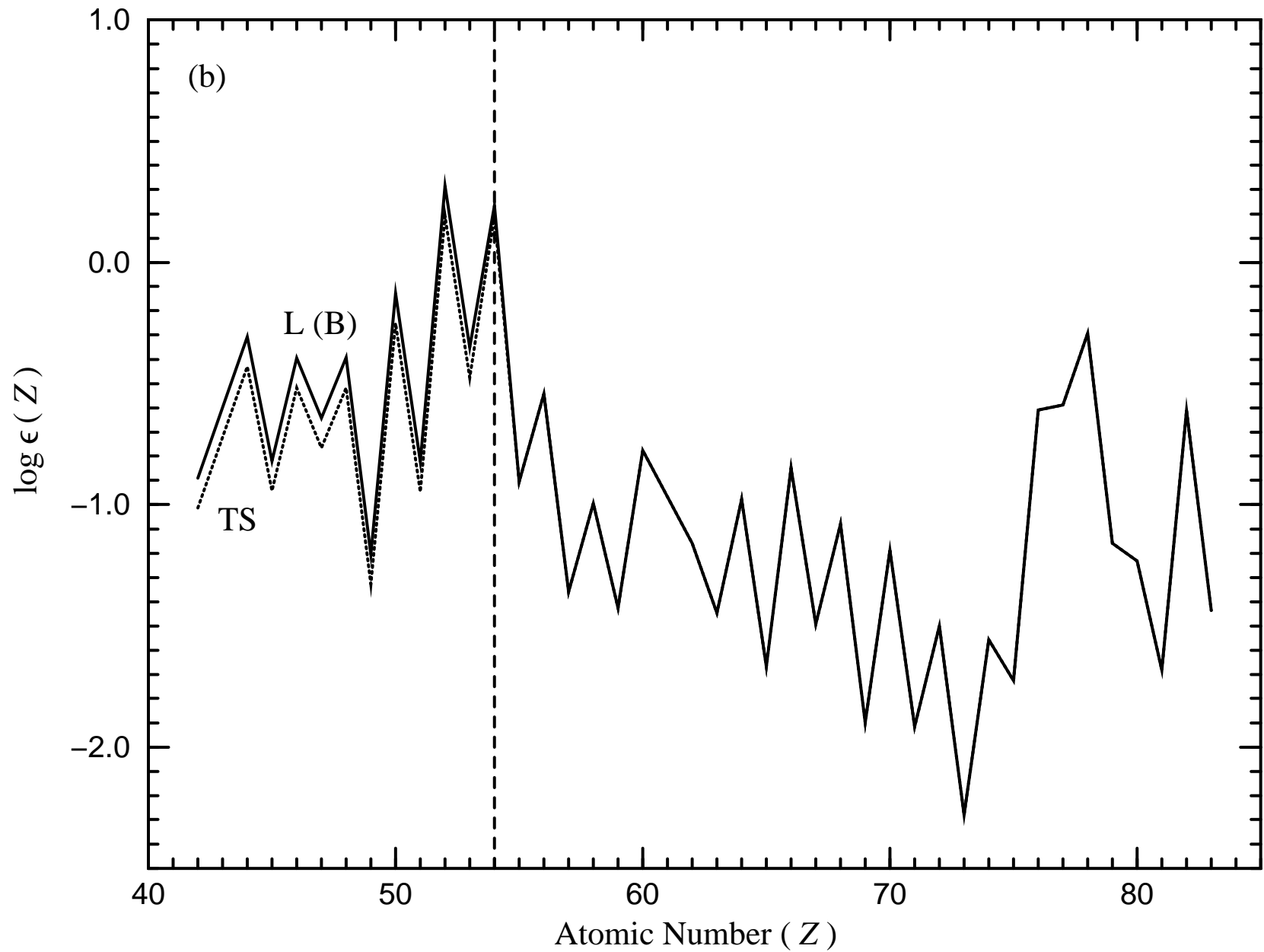
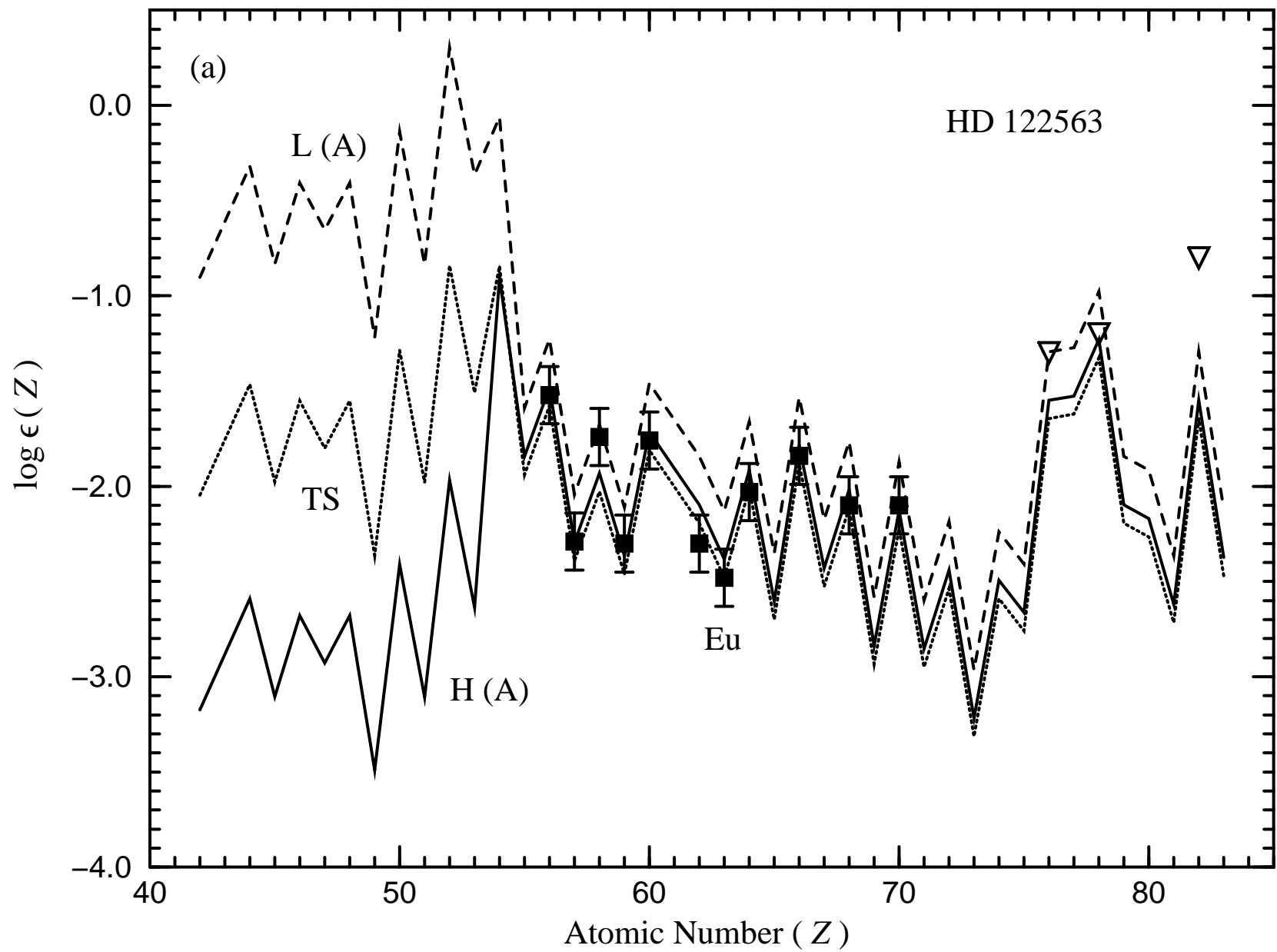
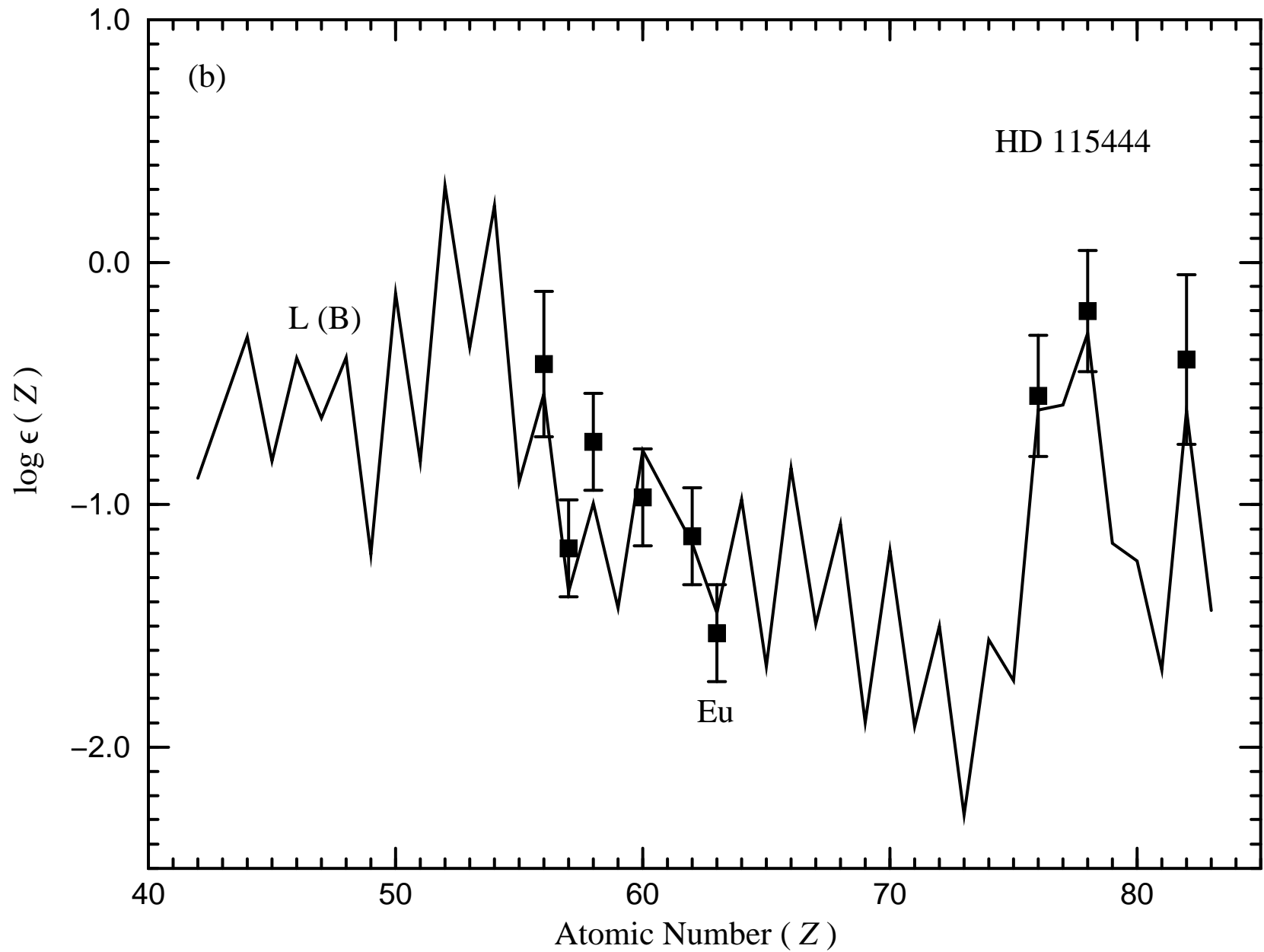
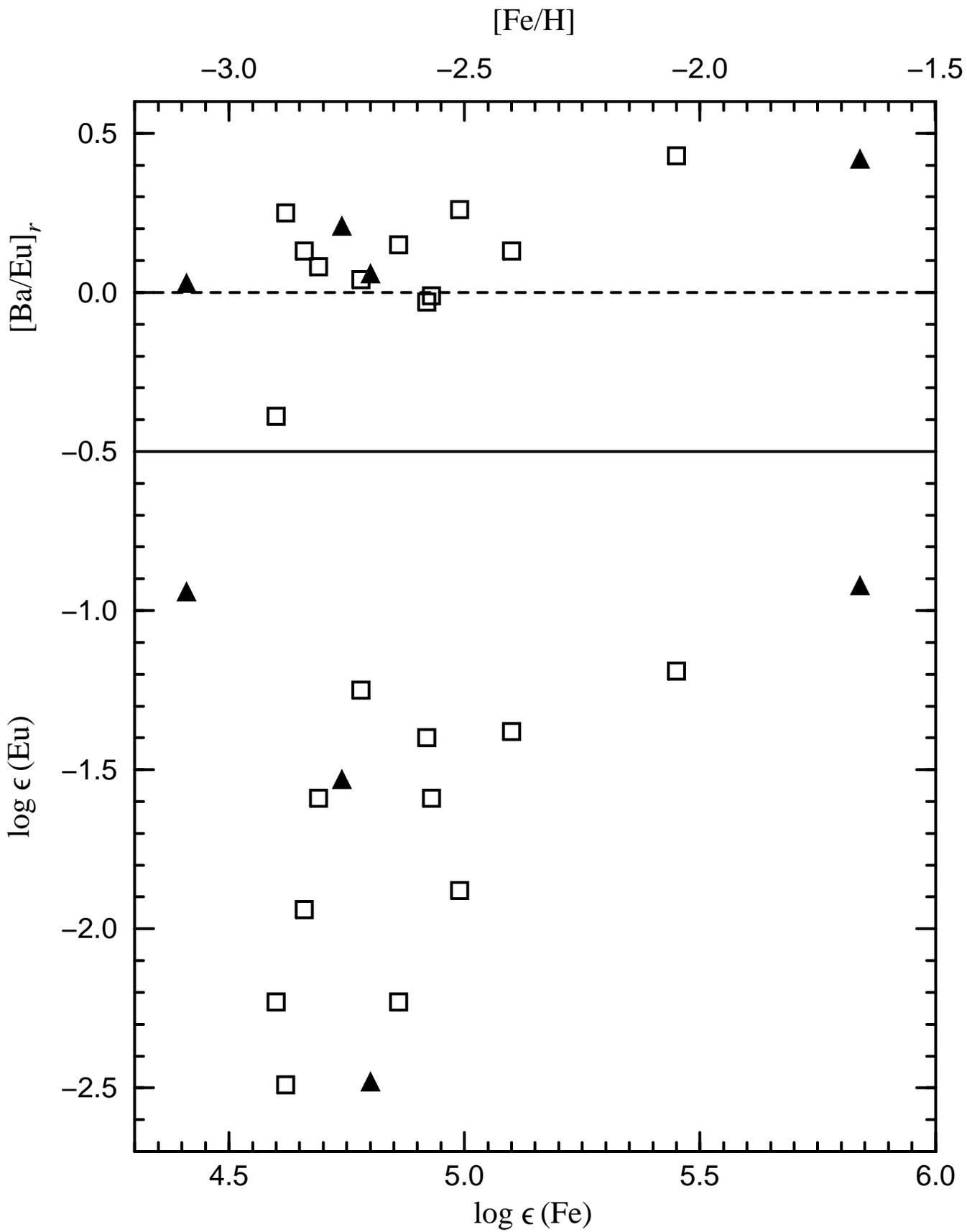
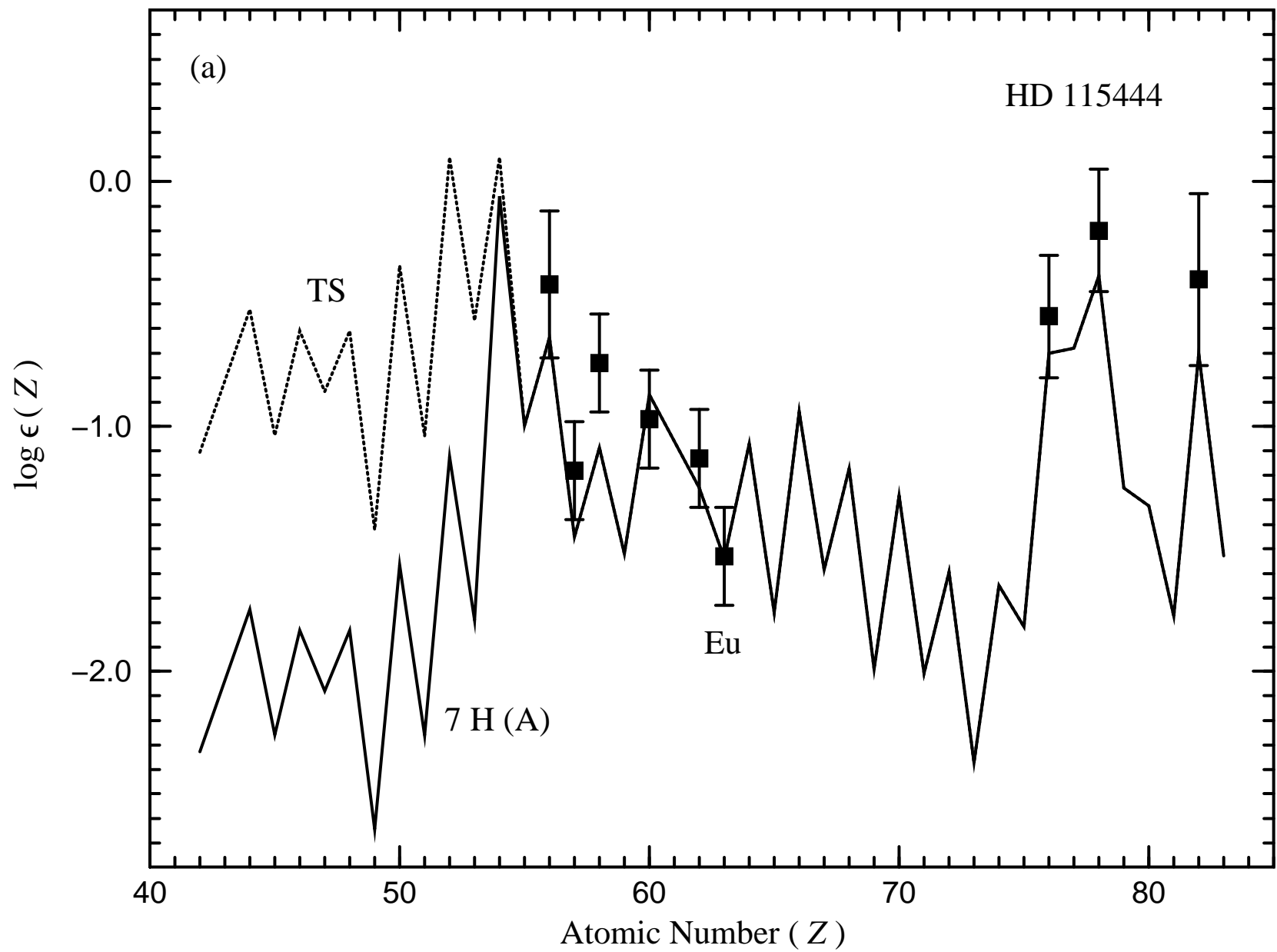
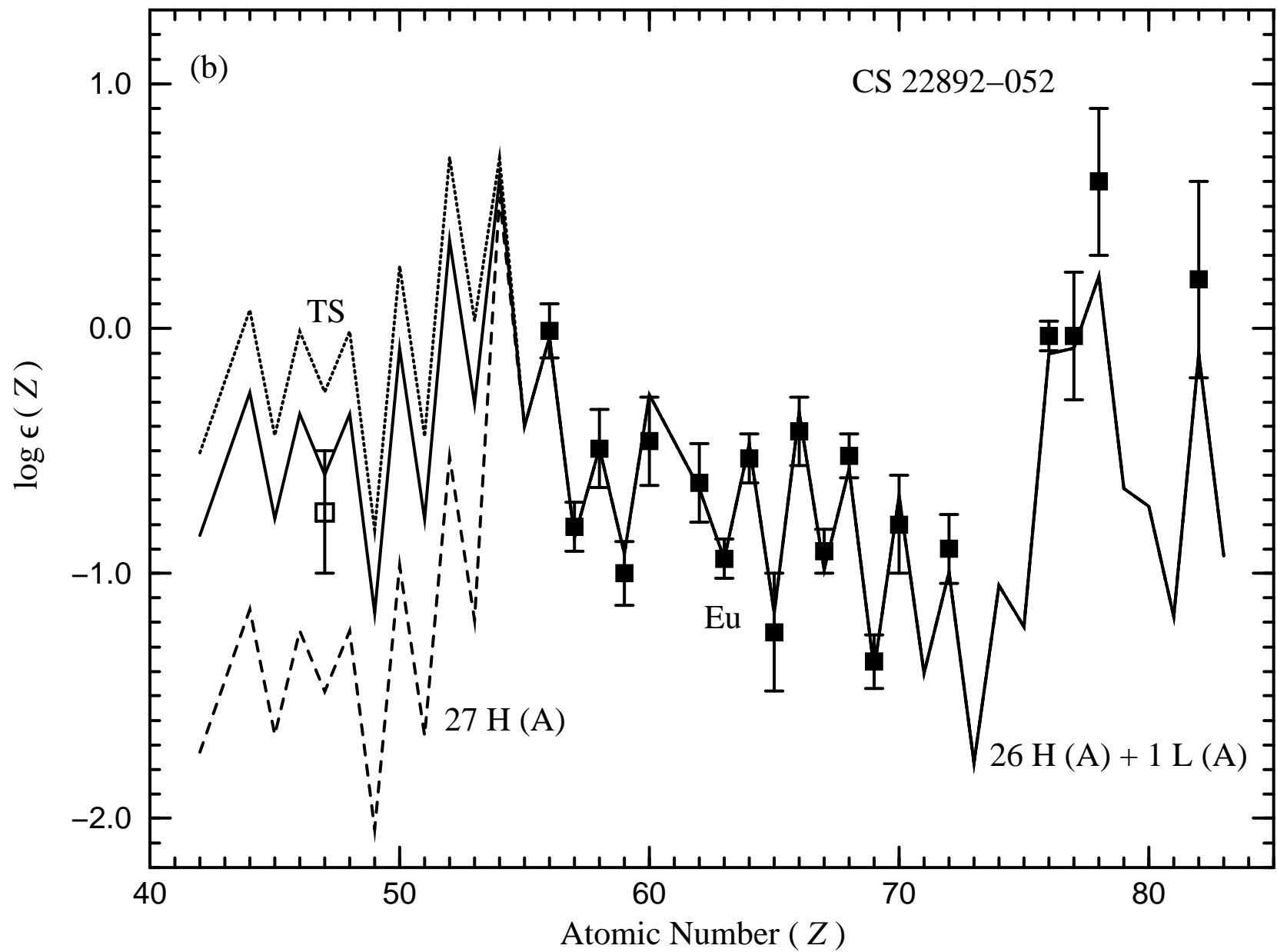
\end{document}